\documentclass[journal,twoside,web]{ieeecolor}
\usepackage{generic}
\usepackage{cite}
\usepackage{algorithmic}
\usepackage{graphicx}
\usepackage{algorithm,algorithmic}
\usepackage{hyperref}
\usepackage{textcomp}

\usepackage{url}
\expandafter\let\csname equation*\endcsname\relax
\expandafter\let\csname endequation*\endcsname\relax 
\usepackage{multirow}
\usepackage{amsmath}
\usepackage{amssymb}
\usepackage{setspace}
\usepackage[czech,english]{babel}
\usepackage{mathrsfs}
\usepackage{mathtools}
\usepackage{physics}
\usepackage{float}
\usepackage{tensor}
\usepackage{tocloft}
\usepackage{mathtools}

\def\vec#1{{\boldsymbol{#1}}}
\def\<{\langle}
\def\>{\rangle}

\def\set#1{{\sf #1}}

\def\map#1{{\mathcal{#1}}}
\def\hil#1{{\mathscr{#1}}}

\def\Cmplx{\mathbb C}

\def\Tr{\operatorname{Tr}}

\def\diag{\operatorname{diag}}
\def\Re{\operatorname{Re}}
\def\Im{\operatorname{Im}}

\newtheorem{theorem}{Theorem}[section]
\newtheorem{definition}{Definition}[section]

\renewcommand{\b}{\beta}

\def\BibTeX{{\rm B\kern-.05em{\sc i\kern-.025em b}\kern-.08em
    T\kern-.1667em\lower.7ex\hbox{E}\kern-.125emX}}
\begin{document}
\bstctlcite{BSTcontrol}
\title{Consensus And Information Redistribution In Quantum Networks}
\author{J. Novotn\'y, D. \v St\v erba, I. Jex, H.-H. Kowalski, and A. Abedi
\thanks{This article was written as part of the Qu-Gov project, which was commissioned by the German Federal Ministry of Finance. J. N., D. Š. and I. J. have been supported by the Grant Agency of the Czech Republic GA ČR under Grant No. 26-23973S. This work was supported by the Grant Agency of the Czech Technical University in Prague, grant No. SGS25/165/OHK4/3T/14. The funders played no role in study design, analysis and interpretation of data, or the writing of this manuscript.}
\thanks{J. Novotn\'y, D. \v St\v erba, and I. Jex is with Department of Physics, Faculty of Nuclear Sciences and Physical Engineering, Czech Technical University in Prague, B\v rehov\'a 7, 115 19 Praha 1 - Star\'e M\v{e}sto, Czech Republic (e-mail: jaroslav.novotny@fjfi.cvut.cz).}
\thanks{H.-H. Kowalski and A. Abedi is with Bundesdruckerei GmbH. Kommandantenstr. 18, 10969 Berlin, Germany (e-mail: jaroslav.novotny@fjfi.cvut.cz).}}

\maketitle

\begin{abstract}
Consensus algorithms are mechanisms driving nodes of a network to agreement. In quantum networks they also redistribute information encoded in the global quantum state, allowing global properties to be inferred from local measurements without reconstructing the global state. Motivated by this, we investigate decentralized quantum consensus in qudit networks governed by discrete random unitary dynamics. We characterize the analytic asymptotic dynamics, identify global information preserved by the evolution, and derive conditions for which different notions of quantum consensus emerge. A general theorem is derived that establishes a unified framework for extracting globally encoded information from local measurements beyond consensus states. First, we analyze CNOT networks and determine which global information can be recovered by individual nodes and which requires joint measurements on cooperating nodes. The CNOT model is well understood and presents a case with entanglement creation – a characteristic quantum feature of networks worth analyzing in detail. Second, we identify the complete class of two-qubit interactions that asymptotically achieve consensus on the average number of excitations, thereby establishing a quantum analogue of classical average consensus. We further show that joint measurements on $k$ cooperating nodes recover exactly the first $k$ moments of the excitation-number distribution.
\end{abstract}


\section{Introduction}
\label{sec:introduction}

\IEEEPARstart{N}{etworks} consisting of many interconnected nodes provide a natural framework for distributed technological, informational, and physical systems \cite{Newman2003,BarabasiAlbert2002}. In such systems, the absence of a central coordinator raises a fundamental question: how can autonomous nodes reach agreement using only local interactions? Consensus algorithms address this problem by enabling nodes to iteratively exchange information and converge to a common value \cite{Lynch1996,Ren2008,Olfati2007}. They underpin numerous applications, ranging from distributed estimation and optimization to cooperative control and sensor networks \cite{Kar2013,Xiao2007,Nedic2009,Olfati2007}.

Classical consensus protocols describe discrete- or continuous-time dynamics in which each node updates its state based solely on information received from neighboring nodes. Consensus is achieved when all node states asymptotically converge to the same value. A particularly important example is average consensus \cite{Olfati2004,Boyd2006}, where all nodes converge to the average of their initial states. This behavior relies on a conservation law: the total sum of node states remains invariant throughout the evolution, allowing each node to recover a global property using only local communication.

Many consensus protocols are implemented through gossip algorithms, where neighboring nodes interact repeatedly in a decentralized and asynchronous manner \cite{Kempe2003,Boyd2006}. Despite relying only on simple local exchanges, such protocols effectively redistribute global information across the network, demonstrating how collective agreement can emerge without centralized control.



The rapid development of quantum technologies, including quantum machine learning \cite{Cong2019,Henderson2020}, quantum sensing \cite{Proctor2018,Zhang2021,Qian2021}, and distributed quantum computing \cite{Broadbent2008,Denchev2008,VanMeter2016}, has generated growing interest in quantum networks. As in classical distributed systems, a central challenge is to design consensus protocols that enable network nodes to reach agreement using only local interactions.


In contrast to the classical setting, both global and local network states are described by density operators, interactions may generate quantum correlations, and information available to individual nodes is obtained from local measurements. These features require new notions of consensus consistent with quantum state dynamics. A systematic framework was introduced in \cite{Mazzarella2015}, where several forms of quantum consensus and their hierarchy were established.

\begin{figure}[htbp]
	\centering
	\includegraphics[width=0.5\textwidth]{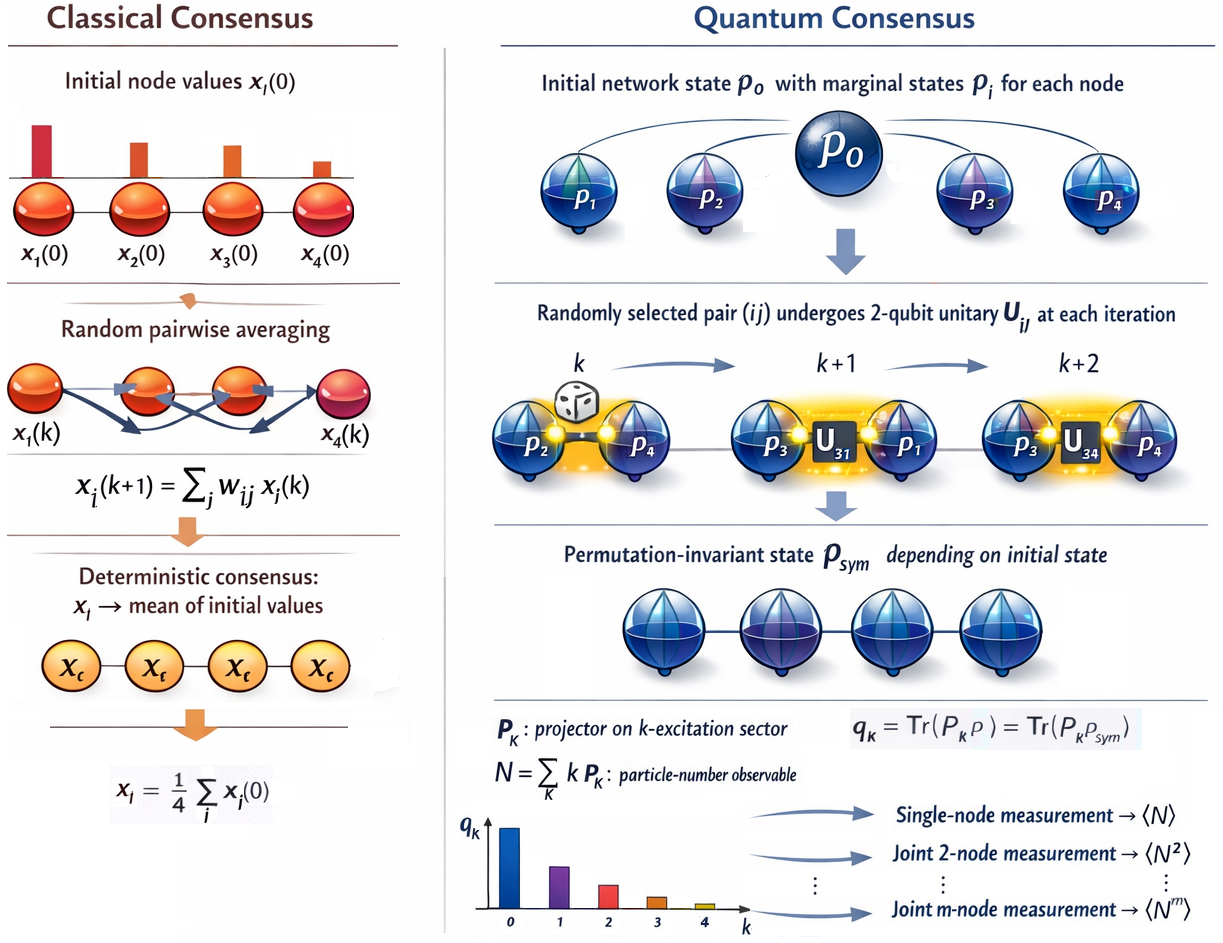}
	\caption{Graphical comparison of classical and quantum average consensus. The classical protocol iteratively converges to the average of the initial node values, while the quantum protocol uses random unitary operations to reach a permutation-symmetric state. Local measurements estimate the average excitation, while collective measurements on $m$ qubits provide access to moments of the excitation-number distribution up to order $m$.}
	\label{fig:classical_versus_quantum_average_consensus}
\end{figure}

One approach to quantum consensus replaces classical gossip steps with random local unitary interactions acting between subsystems of a quantum network without global coordination. A graphical comparison of differences between consensus in classical and quantum networks is illustrated in Fig. \ref{fig:classical_versus_quantum_average_consensus}.
Different protocols based on switching interactions, in continuous \cite{Ticozzi2014,Petersen2016,Petersen2023,Jafarizadeh2016,Jafarizadeh2017} and discrete-time case \cite{Mazzarella2015,Jafarizadeh-discrete2016}, were proposed and investigated including their convergence rates. In \cite{Ticozzi2016}, a consensus protocol beyond the gossip-type algorithms were introduced capable to engineer the network state into symmetric pure Dicke states including GHZ and Werner states. A different type of stochastic consensus protocols with feedback control was designed in \cite{Takeuchi2016} in order to preserve purity of the initial network state.

In this work, we investigate quantum consensus in qudit networks driven by discrete random unitary operations. Our motivation is fourfold. First, this framework provides a natural generalization of decentralized consensus algorithms beyond switching interactions. By allowing general unitary interactions between qudits, it substantially enlarges the class of achievable asymptotic states and consequently leads to a richer variety of consensus states. Second, these protocols are inherently iterative, with consensus emerging asymptotically. The long-time dynamics of random unitary networks can be characterized analytically through the corresponding attractor equations \cite{ruo1,ruo2}, providing a powerful tool for determining their asymptotic behavior. Third, compared with continuous-time Lindblad dynamics, discrete random unitary evolution is conceptually simpler to realize, as it can be implemented through unitary evolution on an enlarged Hilbert space with ancillary systems \cite{Chuang2002}. This discrete formulation is therefore particularly well suited for implementation on current quantum computing platforms, such as IBM Quantum processors. Finally, the proposed framework naturally extends to gossip-type consensus algorithms, such as those introduced in \cite{Mazzarella2015}, as well as to continuous-time formulations based on unitary Lindblad operators \cite{Petersen2016}.

Our primary objective is to investigate consensus algorithms as a mechanism for preserving global information encoded in the state of a quantum network and uniformly redistributing this information among its nodes, where it can be independently extracted through local measurements. Because the global state of a quantum network contains substantially richer information than its classical counterpart, this aspect of consensus becomes particularly significant. The central question is therefore not only whether the network reaches consensus, but also what information the nodes ultimately agree upon. This question is of practical importance because full quantum state tomography rapidly becomes infeasible as the network size increases. Consensus algorithms offer an alternative by redistributing global information to local nodes, allowing it to be recovered from local measurements without reconstructing the entire network state. 

An important first step in this direction was taken in \cite{Mazzarella2015}, where the authors introduced the notion of globally encoded information redistributed to local network nodes and characterized it in terms of expectation values of global and local observables. In the present work, using the attractor framework for random unitary operations, we identify the global information preserved throughout the consensus process, determine under which conditions different types of consensus are attained, and formulate a general theorem that characterizes the information accessible to one or more nodes through measurements of arbitrary local observables once the network has reached its asymptotic regime. The significance of this theorem extends well beyond the characterization of consensus states. It provides a general framework for determining which global properties of a quantum state can be recovered from local measurements. We discuss its applicability in detail and present a systematic procedure for identifying the global information that can be extracted either by a single node or by several cooperating nodes performing joint measurements.

Finally, we apply the developed theory to two representative examples. First, we investigate CNOT networks with controlled-NOT interactions, whose attractor structure is known \cite{Novotny2011}, and determine which global information is accessible from measurements on a single node and which requires joint measurements on two nodes. In the second case, we employ the theory for the reverse task. In particular, we identify the complete class of two-qubit interactions that achieve consensus on the average number of excitations in the network. This form of quantum consensus can be regarded as a direct analogue of classical average consensus in a decentralized setting. We further show that allowing $k$ nodes to cooperate enables the recovery of exactly the first $k$ moments of the probability distribution of the total number of excitations in the network.

The paper is organized as follows. In section \ref{sec:setting the scene}, the mathematical formalism of qudit networks is provided including various definitions of quantum consensus and the concept of redistribution of global information into local nodes. Section \ref{sec:Random unitary evolution towards consensus} is devoted to introduction of random unitary dynamics for qudit networks. Their asymptotic theory is summarized and theorems for consensus identification and information redistribution within networks are derived. In section \ref{chap:CNOTs} we analyze consensus formation in networks with CNOT interaction. QAC for the total number of excitations is explored in section \ref{sec:total number of excitation}. Summary and conclusion is provided in section \ref{sec:conclusion}.

\section{Quantum consensus - setting the scene}
\label{sec:setting the scene}

In the quantum setting, information is stored within quantum states. To formalize this framework, throughout this paper we consider a quantum system associated with a finite-dimensional Hilbert space $\mathcal{H}$. When the Hilbert space is two-dimensional, the system is called a \emph{qubit}; more generally, a system defined on a $d$-dimensional Hilbert space is termed a \emph{qudit}. The state of a qudit is described by a density operator $\rho: \mathcal{H} \rightarrow \mathcal{H}$, which is a positive semidefinite operator with unit trace. This operator fully characterizes the information accessible through quantum measurements. The density operator determines the expectation value of any observable $\sigma = \sigma^\dagger : \mathcal{H} \rightarrow \mathcal{H}$
(a Hermitian operator) via the formula $\langle \sigma \rangle_{\rho} = \mathrm{Tr}(\sigma\rho)$.
Conversely, knowledge of the expectation values of all observables allows for the reconstruction of the quantum state. Therefore, the informational content of a quantum state is fully captured by the mean values of all observables.

We further consider a system composed of $n$ spatially separated distinguishable qudit subsystems. The $i$-th qudit is associated with the Hilbert space $\hil H_i$ ($i\in \{1,2,\ldots, n\}$) of dimension $d$. Hence all these Hilbert spaces are mutually isomorphic $\hil H_i \cong \hil H$. This system can be represented as a quantum network, where each node possesses or corresponds to one qudit. The interpretation of the links between nodes will be discussed in Section \ref{sec:Random unitary evolution towards consensus}. The Hilbert space of the entire network is given by $\hil H^n = \otimes_i \hil H_i \cong \hil H^{\otimes n}$, and the joint (global) quantum state is represented by a density operator $\rho:\hil H^n \rightarrow \hil H^n$. The local reduced state $\rho_i: \hil H_i \rightarrow \hil H_i$ of the $i$-th qudit is obtained by tracing out all the other qudits, i.e.
\begin{equation}
    \rho_i=\Tr_{\otimes_{j \neq i} \hil H_j} \rho.
\end{equation}
The consistency between the expected value of any local observable $\sigma$ and corresponding measurement on the whole system implies 
\begin{equation}
    \langle \sigma \rangle_{\rho_i} = \Tr_{\hil H_i}(\sigma\rho_i) = \Tr_{\otimes_i \hil H_i}\left(\sigma^{(i)}\rho\right) = \langle \sigma^{(i)} \rangle_{\rho},
\end{equation}
where we denoted 
$$\sigma^{(i)}=I^{\otimes (i-1)}\otimes \sigma \otimes I^{\otimes (n-i)}$$
with $A^{\otimes k}$ being the tensor product $A\otimes A \otimes \ldots \otimes A$ of $k$ factors and $I$ the identity operator acting on the one-qudit Hilbert space.
Since each node has access only to its local qudit subsystem, it possesses limited knowledge of the global state. As a result, the information encoded in the overall quantum state is typically distributed asymmetrically across the network. The central quantum information task is to develop a protocol or a dynamical process that drives the network toward a state in which the information is kept and evenly distributed among the nodes. When such a state is reached, the network is said to have achieved consensus. The specific form of consensus depends on the agreement criteria required whose detailed classification is provided in \cite{Mazzarella2015}. In the following, we summarize the forms of quantum consensus that are relevant for this work. We start with the weakest form of consensus. 


\begin{definition}
    The $\sigma$-expectation consensus ($\sigma EC$) occurs for the state $\rho$ if it follows for a given observable $\sigma$ and any pair of nodes $i$ and $j$
    \begin{equation}
        \label{sigma_expactation_consensus}
        \Tr(\sigma^{(i)}\rho) = \Tr(\sigma^{(j)}\rho)=\Tr(\sigma \rho_i) = \Tr(\sigma \rho_j).
    \end{equation}
\end{definition} 
Simply, $\sigma EC$ occurs in a quantum network if all its nodes measure the same mean value of one-qudit observable $\sigma$, i.e. they share the same local information $\<\sigma\>$. Apparently, it does not matter in which order they measure this local information, because local measurements on one subsystem can not impact statistics of local measurements on another subsystem. 

We receive a stronger form of quantum consensus if $\sigma EC$ is achieved for an arbitrary one-qudit observable $\sigma$. 
\begin{definition}
    The reduced state consensus (RSC) occurs for the state $\rho$ if all its one-qudit reduced states are the same, i.e. for any pair of indices $i$ and $j$
    \begin{equation}
        \label{reduced_state_consensus}
        \rho_i =\rho_j.
    \end{equation}
\end{definition} 
The quantum network is in the RSC, if all nodes possess the same local one-qudit state. An even stronger form of consensus is provided in the next definition.
\begin{definition}
    The symmetric state consensus (SSC) occurs for the network state $\rho$ if it is symmetric under an arbitrary permutation, i.e., for any permutation $\pi$ of the order of all network nodes the state $\rho$ follows
    \begin{equation}
        \label{symmetric_state_consensus}
        U_{\pi}\rho U_{\pi}^{\dagger}=\rho,
    \end{equation}
    where the unitary operator $U_{\pi}$ is uniquely defined as
    $$
    U_{\pi} X_1\otimes X_2 \otimes \ldots X_n U_{\pi}^{\dagger}= X_{\pi(1)}\otimes X_{\pi(2)}\ldots X_{\pi(n)}
    $$
    for any $n$ operators $X_i$ acting on $\hil H$.
\end{definition} 
This implies that network subsystems of equal size are all in the same state. In such a case, if two or more nodes cooperate together and measure the mean value of some two- or more-qudit observable, they will obtain the same information, irrespective of the choice of nodes. 

The redistribution of information arises from all prior processes that have occurred within the network, whether they were deliberately implemented or arose from unavoidable interactions with the surrounding environment. As a result, the network is not typically in any consensus state mentioned above. 

Our main goal is to design an algorithm (network dynamics) that would bring any network state to a desired type of consensus with certain distributed information. Such an algorithm should preserve a certain global information about the state of the whole network and redistribute it among all nodes in a way that it can be read locally and independently by an arbitrary network node. In particular, assume that the dynamics of the network evolves from an arbitrary initial state $\rho(0)$ to a state $\rho(t)$ at time $t$. 
\begin{definition}
    We say that the network asymptotically reaches $\Sigma-average$ $\sigma EC$ if it asymptotically achieves $\sigma EC$ and the mean value of a global observable $\Sigma:\hil H^n \rightarrow \hil H^n$ satisfies 
    \begin{equation}
        \label{sigma_average_expactation_consensus}
       \<\Sigma \>_{\rho(0)}=\<\Sigma \>_{\rho(t)} = \lim_{t \rightarrow \infty}\<\sigma\>_{\rho_i(t)}
    \end{equation}
    for an arbitrary node $i$ and any initial network state $\rho(0)$.
\end{definition} 
In other words, $\Sigma-average$ $\sigma EC$ is achieved asymptotically if the network dynamics preserves global information $\<\Sigma\>_{\rho(0)}$, i.e. $\Sigma$ is its integral of motion, and redistributes it along all nodes so that it can be read locally by measuring the mean value of the local observable $\sigma$.

\section{Random unitary evolution toward consensus}
\label{sec:Random unitary evolution towards consensus}

In this section, we introduce a class of quantum dynamics that provides a suitable approach for driving a network toward specific types of consensus. We develop a general framework to determine which forms of consensus can be achieved under a given dynamics from this class, and to characterize what information is preserved and how it can be extracted from one or more nodes.

Before proceeding, a natural question arises: can existing classical algorithms be leveraged to realize at least some of the previously defined consensus types in a quantum network? For certain types of additive distributed information, classical consensus algorithms may indeed offer a viable strategy. However, this approach requires extensive measurements at each node to extract classical data, which is then processed using classical methods, making it highly resource-intensive. A more detailed discussion is provided in Section \ref{sec:total number of excitation}.

In a fully quantum consensus algorithm, quantum interactions among different qudit subsystems take over the role of classical communication in classical algorithms. Similarly to the classical case, we assume that due to spatial separation, qudits interact only pairwise with their neighbors. This provides a certain interaction pattern that can be captured by links in the network. These links simply encode which pairs of qudits eventually undergo bipartite interaction. They can be directed or undirected, based on whether interaction is symmetric or not, but otherwise the form of interaction is not specified yet. 

We consider a symmetric redistribution of information across local nodes, meaning that each node obtains identical information. The objective is to drive the entire network toward an asymptotic (not necessarily stationary) state characterized by equal expectation values of selected local observables at every node. Since such a configuration cannot be realized through unitary evolution alone, it requires the inclusion of auxiliary (ancillary) quantum systems. Consequently, the network dynamics must be inherently open, allowing for partial information loss through the suppression of certain quantum correlations associated with the system’s initial state.

\begin{figure}[htbp]
	\centering
	\includegraphics[width=0.4\textwidth]{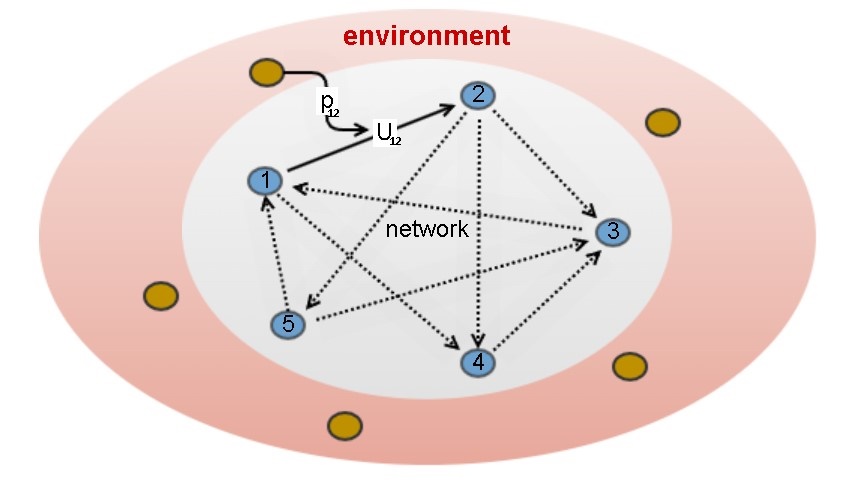}
	\caption{Example of a quantum network with 5 qudits driven by random unitary interaction.}
	\label{fig:open_network}
\end{figure}

A simple model that simultaneously ensures the local nature of interactions and the openness of the overall dynamics is a network governed by discrete, time-resolved, bipartite random unitary interactions. Each link in the network represents an eventual interaction of the connected nodes. Links may be oriented, representing non-symmetric interactions, and link weights represent probabilities with which the corresponding interaction occurs within one step of evolution. In summary, the vertices of the network $\set G(\set V,\set E)$ represent qudits of the network, the links represent ordered pairs of qudits that are allowed to interact unitarily and weights of links corresponds to probabilities with which they two qudits interact. The unitary interaction is assumed to be given by the same two-qudit unitary gate $U$. If it acts on the ordered pair of qudits $e \in \set E$ we denote it by $U_e$. We assume that in every step of the evolution the interaction between qudits occurs randomly, i.e. each edge $e$ of the graph $\set G(\set V,\set E)$ is weighted by the probability $p_e$ with which the network undergoes interaction $U_e$. Due to the uncontrolled randomness of individual pairwise interactions between qudits, the overall one-step evolution of the network is given by 
\begin{equation}
\label{random evolution of network}
    \rho(t+1)= \map P(\rho(t))= \sum_{e \in \set E} p_e U_e \rho(t) U_e^{\dagger}.
\end{equation}
The propagator $\map P$ is a random unitary operation, a quantum operation with Kraus operators $\sqrt{p_e}U_e$ \cite{Chuang2002}. It is uniquely prescribed by the two-qudit unitary gate and the interaction network graph $\set G(\set V,\set E)$, whose example is depicted in Fig. \ref{fig:open_network}. 

After $t$ steps of evolution, the state of the network originally initiated in the state $\rho(0)$ is
\begin{equation}
    \label{network evolution}
    \rho(t) = \map P^{t}(\rho(0)).
\end{equation}
It describes an open-system evolution that progressively narrows the range of accessible states, ultimately reaching its asymptotic regime $\rho_{as}(t)=\rho(t \gg 1)$. The characteristics of this asymptotic evolution are governed exclusively by the structure of the bipartite interaction $U$ and the topology of the interaction network $\set G(\set V,\set E)$. The key question is how to design these elements in order to steer the network toward a desired form of quantum consensus. 


\subsection{Asymptotic evolution of networks generated by RUO}
\label{chap:asymptotic theory}

In this section, we review the asymptotic theory of random unitary operations; for further details, see \cite{ruo1,ruo2}. Building on this foundation, we develop a framework for identifying various forms of consensus in quantum networks.

The analysis of the random unitary evolution in (\ref{random evolution of network}), compared to a closed unitary evolution, is typically more complex because the generator $\map P$ is not, in general, normal or even Hermitian. Consequently, it cannot be diagonalized into an orthonormal set of eigenmatrices in general. However, when studying consensus formation, our focus is solely on the asymptotic regime of this evolution, which can always be diagonalized. The asymptotic dynamics of evolution (\ref{network evolution}) takes place in the so-called attractor space, defined as the span of all solutions $X$ of the attractor equations
\begin{equation}
    \label{eq_attractors_general}
    U_{e} X = \lambda X U_e,
\end{equation}
simultaneously fulfilled for all $e \in \set{E}$ and an arbitrary $\lambda$ with $|\lambda|=1$. Solutions that satisfy all equations (\ref{eq_attractors_general}) for some $|\lambda|=1$ are called attractors associated with the eigenvalue $\lambda$, and together they form the asymptotic spectrum $\sigma_{as}$. The asymptotic spectrum is always nonempty, as it contains eigenvalue one associated with the attractor proportional to the maximally mixed state. It is important to stress that the attractor equations (\ref{eq_attractors_general}) imply that the probabilities with which the unitaries enter evolution have no impact on the resulting asymptotic dynamics, provided there are different from zero. Once an orthonormal basis $\{X_{\lambda,i}\}$ of all solutions $X$ of equation (\ref{eq_attractors_general}) for all admissible $|\lambda|=1$ is known, the general asymptotic evolution of the initial state $\rho(0)$ is given as
\begin{equation}
\label{eq_asymptotic_evolution_general}
\rho_{as}(t)=\rho(t \gg 1) = \sum_{\lambda \in \sigma_{as},i} \lambda^t \Tr\{\rho(0)X_{\lambda,i}^{\dagger}\} X_{\lambda,i},
\end{equation}
where $i$ accounts for the possible degeneracy of solutions $X_{\lambda,i}$ of (\ref{eq_attractors_general}) for a given $\lambda$.

In the Heisenberg picture, the one step of evolution of an observable $A(t)$ is given by the random unitary operation
\begin{equation}
    \label{heisenberg_picture_ruo}
    A(t+1)=\sum_{e \in \set{E}} p_e U_e^{\dagger}A(t)U_e.
\end{equation}
Its attractor space coincides with the attractor space of evolution in the Schr\"odinger picture, and 
hence the asymptotic evolution of observable $A(0)$ in the Heisenberg picture reads 
\begin{equation}
\label{eq_asymptotic_evolution_general_heisenberg}
A(t \gg 1) = \sum_{\lambda \in \sigma_{as},i} \overline{\lambda}^t \Tr\{A(0)X_{\lambda,i}^{\dagger}\} X_{\lambda,i}.
\end{equation}
In view of equations (\ref{eq_asymptotic_evolution_general}) and (\ref{eq_asymptotic_evolution_general_heisenberg}), the attractors encode the process's memory of the initial state. From the complete set of attractors, one can construct all the constants of motion of the process that fully describe its asymptotic evolution \cite{Novotny2023}. These are generally time-dependent observables in the Heisenberg picture (\ref{heisenberg_picture_ruo}), whose expectation values are conserved along every trajectory. Note also that fixed points (attractors associated with $\lambda=1$) determine the set of stationary states and also the set of integrals of motion (charges); steady constants of motion.

\subsection{Quantum Consensus Characterized by Attractors}
\label{chap:Quantum Consensus Characterized by Attractors}

Equipped with the asymptotic theory of RUOs, we can now examine whether a quantum network with random unitary interactions evolves toward a specific type of quantum consensus state. The following theorem outlines the algebraic conditions corresponding to each type of introduced consensus.
\begin{theorem}
\label{theorem:consensus check}
A quantum network with random unitary dynamics (\ref{random evolution of network}) evolves any initial state $\rho(0)$ toward
\begin{itemize}
    \item[1.] the $\sigma$-expactation consensus iff all its attractors $X_{\lambda,k}$ satisfy
    \begin{equation}
        \label{sigma-expactation consensus attractors}
        \Tr(\sigma^{(i)}X_{\lambda,k}) = \Tr(\sigma^{(j)}X_{\lambda,k})
    \end{equation}
     for the observable $\sigma$ and any pair of qudits $i$ and $j$.
    \item[2.] the reduced state consensus iff the reduced operators of all its attractors $X_{\lambda,k}$ follow
    \begin{equation}
        \label{reduced state consensus attractors}
        \Tr_{l\neq i} X_{\lambda,k} =   \Tr_{l\neq j} X_{\lambda,k}
    \end{equation}
    for any pair of qudits $i$ and $j$.
    \item[3.] the symmetric state consensus iff all its attractors $X_{\lambda,k}$ satisfy
    \begin{equation}
        \label{symmetric state consensus attractors}
        U_{\pi} X_{\lambda,k} U_{\pi}^{\dagger} = X_{\lambda,k}
    \end{equation}
    for any qudit permutation $\pi$.
\end{itemize}
\end{theorem}
{\bf Proof}:  We prove the first statement; the others are analogous. The sufficient condition is obvious from the form of asymptotic states (\ref{eq_asymptotic_evolution_general}). We focus on the necessary condition. The network approaches $\sigma EC$ if the condition (\ref{sigma_expactation_consensus}) is fulfilled for any asymptotic state (\ref{eq_asymptotic_evolution_general}), which can be rewritten as
\begin{equation}
    \sum_{\lambda\in\sigma_{as},k} \lambda^t \Tr\{X_{\lambda,k}^{\dagger}\rho(0)\}\Tr\left[X_{\lambda,k}\left(\sigma^{(i)}-\sigma^{(j)}\right)\right]=0
\end{equation}
valid for an arbitrary density operator $\rho(0)$ and any pair of qudits $i$ and $j$. Since any operator $Z$ can be expressed as a linear combination of density operators and due to the linearity of this equation with respect to $\rho(0)$, it must hold for an arbitrary operator $Z$ as well. If we choose $Z=X_{\tilde{\lambda},m}$, the orthogonality of the attractor basis $\{X_{\lambda,k}\}$ gives us
\begin{equation*}
    \Tr\left[X_{\tilde{\lambda},m}\left(\sigma^{(i)}-\sigma^{(j)}\right)\right]=0,
\end{equation*}
which proves the statement. $\blacksquare$

One of the main objectives of this work is to investigate the redistribution of global network information to local subsystems and its subsequent reading using local measurements. In particular, it encompasses the following features which we discuss in detail from the point of view of attractors.
\begin{itemize}
    \item 
    For sufficiently large iteration numbers, the network state approaches its asymptotic subspace, which always contains the steady maximally mixed state. The asymptotic state is unique iff the attractor space is one-dimensional. In this case, all initial states converge to the maximally mixed state and the dynamics has no memory. In contrast, quantum consensus with information redistribution requires a nontrivial attractor space of dimension at least two, implying at least two linearly independent conserved quantities.
    \item We further assume that preserved information is redistributed to local nodes in the asymptotic regime and can be accessed by measuring the expectation value of a time-independent local observable. Therefore, we focus on the information encoded solely in the expectation values of the integrals of motion. In other words, the network is supposed to reach asymptotically $\Sigma-average$ $\sigma EC$ defined by relation (\ref{sigma_average_expactation_consensus}).
\end{itemize}

The crucial task is to find which global information may be propagated into local subsystems and subsequently read in a given quantum network. The following theorem provides a systematic attractor-based method allowing one to identify which global information is locally accessible in a specific node $j$ of a given network with random unitary evolution (\ref{random evolution of network}).

\begin{theorem}
\label{theorem:local_observable}
Assume a qudit network with random unitary evolution (\ref{random evolution of network}). Let $\{X_{\lambda,i}\}$ be an orthonormal basis of its attractor space with $\{Y_i\}$ denoting the subset of this basis corresponding to eigenvalue one. Then the mean value of a local one-qudit observable $\sigma$ measured at the network node $j$ satisfies for some integral of motion $\Sigma_j$ in the asymptotic evolution
\begin{equation}
    \Tr\{\rho_{as,j}(t) \sigma \}=\Tr\{\rho(0)\Sigma_j\}
\end{equation}
for any initial network state $\rho(0)$ and its corresponding asymptotic state $\rho_{as}(t)$ if and only if
\begin{itemize}
    \item[1.] overlap of $\sigma$ with all reduced non-stationary attractors $X_{\lambda,i}$ of the $j$-th node is equal to zero, i.e.
    \begin{equation}
        \label{zero overlaps}
        \Tr\left\{X_{\lambda,i}^{(j)}\sigma \right\} = 0
    \end{equation}
    for all $X_{\lambda,i}$ with $|\lambda|\neq 1$. The operator $X_{\lambda,i}^{(j)}=\Tr_{k\neq j} X_{\lambda,i}$ is obtained by tracing out all qudits except the $j$-th.
    \item[2.] $\Sigma_j$ is an integral of motion uniquely defined as
    \begin{equation}
    \label{Sigma_general}
    \Sigma=\Sigma_j = \sum_i \overline{\Tr\left\{\sigma Y_i^{(j)}\right\}} Y_i.
\end{equation}
The subscript $j$ in global observable $\Sigma_j$ refers to the node $j$ at which its mean value is obtained by measuring the mean value of the local observable $\sigma$.
\end{itemize}
In addition,  the network asymptotically reaches $\Sigma-average$ $\sigma$-expectation consensus (\ref{sigma_average_expactation_consensus}) for all initial states $\rho(0)$ if and only if conditions (\ref{zero overlaps}) are fulfilled for all nodes of the network and the integral of motion $\Sigma=\Sigma_j$ does not depend on the selected node $j$, i.e. $\Sigma_j=\Sigma_k$ for any pair of nodes $j$ and $k$.
\end{theorem}
{\bf Proof}: In general, attractor spaces of a random unitary evolution in both, Schr\"{o}dinger and Heisenberg, pictures coincide. Hence, any integral of motion has the form $\Sigma = \sum_i \gamma_i Y_i = \sum_i \Tr\{Y_i^{\dagger}\Sigma \} Y_i $. Similarly, any asymptotic state $\rho_{as}(t)$ initiated in the state $\rho(0)$ can be expressed as $
\rho_{as}(t) = \sum_{\lambda \in \sigma_{as},i}\lambda^t X_{\lambda,i}\Tr\{X_{\lambda,i}^{\dagger}\rho(0) \}$ and thus its corresponding reduced state of the $j$-th node is  equal to $\rho_{as,j}(t) = \sum_{\lambda \in \sigma_{as},i}\lambda^t X_{\lambda,i}^{(j)}\Tr\{X_{\lambda,i}^{\dagger}\rho(0) \}$.  The mean value for the global integral of motion then reads as follows
\begin{equation}
    \<\Sigma\>_{\rho(0)}= \Tr\left\{\Sigma^{\dagger} \rho(0)\right\} = \sum_i \overline{\gamma}_i \Tr\left\{Y_i^{\dagger} \rho(0)\right\}.
\end{equation}
Analogously, the mean value of the local observable $\sigma$ for the reduced state $\rho_{as,j}(t)$ is given as
\begin{equation}
    \< \sigma\>_{\rho_{as,j}(t)} =   \sum_{\lambda \in \sigma_{as},i}\lambda^t \Tr\left\{\sigma X_{\lambda,i}^{(j)} \right\} \Tr\left\{X_{\lambda,i}^{\dagger} \rho(0) \right\}.
\end{equation}
The equality of both mean values, i.e. $\<\Sigma\>_{\rho_0}=\< \sigma\>_{\rho_{as,j}(t)}$, implies
\begin{eqnarray}
\label{mean values equality}
    &&\sum_{\lambda \in \sigma_{as},\lambda\neq 1,i}\lambda^t\Tr\left\{X_{\lambda,i}^{\dagger}\rho(0)\right\}\Tr\left\{\sigma X_{\lambda,i}^{(j)} \right\}  \nonumber \\&+&\sum_i \Tr\left\{Y_i^{\dagger} \rho(0)\right\} \left[\overline{\gamma}_i - \Tr\left\{\sigma Y_i^{(j)} \right\} \right]=0,
\end{eqnarray}
where we decompose the sum over all attractors into a contribution from those with eigenvalue 1 and a contribution from all remaining attractors. Since (\ref{mean values equality}) must be fulfilled for any initial state $\rho(0)$, due to linearity, it must be valid for any operator $Z$ and thus also for $Z=Y_k$. The orthogonality of attractors $X_{\lambda,k}$ (and consequently $Y_k$) implies finally
\begin{equation}
    \gamma_k= \overline{\Tr\left\{\sigma Y_k^{(j)}\right\}}.
\end{equation}
Hence $\Sigma=\Sigma_j$ and (\ref{mean values equality}) results in 
\begin{eqnarray}
\label{mean values equality 2}
    \sum_{\lambda \in \sigma_{as},\lambda\neq 1,i}\lambda^t\Tr\left\{X_{\lambda,i}^{\dagger}\rho(0)\right\}\Tr\left\{\sigma X_{\lambda,i}^{(j)} \right\}  \nonumber =0.
\end{eqnarray}
This equation must be valid for any density operator $\rho(0)$. Similarly as in (\ref{mean values equality}), this gives us $\Tr\left\{X_{\lambda,i}^{(j)}\sigma \right\} = 0$ for any non-stationary $X_{\lambda,i}$ from orthonormal basis $\{X_{\lambda,i}\}$.

The last claim of the theorem follows from the fact that $\Sigma$-average $\sigma EC$ consensus occurs iff all nodes read the same information by measuring the mean value of the observable $\sigma$. Therefore, condition (\ref{zero overlaps}) must hold for any node and since they have to measure the same mean value, all integrals of motion $\Sigma_j$ must coincide. $\blacksquare$

Theorem \ref{theorem:local_observable} has two key implications. First, it identifies the conditions under which measuring the expectation value of a local observable $\sigma$ yields information about the global state of the network. Specifically, the local observable must have zero overlap with all corresponding reduced non-stationary attractors of the $j$-th node. This requirement is automatically satisfied when the asymptotic dynamics of the network is stationary, i.e., when its random unitary evolution has no attractors associated with eigenvalues $\lambda \neq 1$. An example of such networks is analyzed in section \ref{chap:CNOTs}. The condition is also satisfied when non-stationary attractors exist but all of their reduced operators vanish, as discussed in section \ref{sec:total number of excitation}. However, in general, this condition must always be verified.

Second, if this condition holds for a chosen local observable $\sigma$, Theorem \ref{theorem:local_observable} determines which specific piece of global information about the network state is accessible to node $j$ through the measurement of its mean value. In particular, the choice of the local observable $\sigma$ uniquely fixes a global observable, time-invariant observable $\Sigma_j$ (\ref{Sigma_general}), whose expectation value is obtained via local measurement. In other words, rather than examining whether a given global observable can be accessed locally and in what manner, Theorem \ref{theorem:local_observable} adopts a reverse viewpoint: one first chooses a local observable and then determines which global observable is accessed through its local measurement at the given node of a quantum network.

Let us discuss the scope of Theorem \ref{theorem:local_observable}, which is considerably broader. First, it applies not only to the case of uniformly distributed information among individual nodes in consensus formation, but also to arbitrary asymmetric redistribution of information across the network. In the latter case, each node reads a different global information represented by a different global observable $\Sigma_j$ defined in (\ref{Sigma_general}), since the coefficients $\Tr\{\sigma Y_i^{(j)}\}$ do not coincide for different nodes in general. By contrast, when these coefficients are equal, the network asymptotically reaches $\Sigma-average$ $\sigma EC$.

Second, Theorem \ref{theorem:local_observable} also applies to subsystems of an arbitrary size, not only to one-qudit subsystems. It will be shown in section \ref{chap:CNOTs} that although some global information can not be read from the measurement on one qudit within the given network, it can still be obtained by mutual action (joint measurement) of two or more qudit nodes. Analogously to the one-qudit case, the global observable whose mean value is obtained by collective measurement of the mean value of $k$-qudit observable $\sigma$ measured in nodes $j_1,j_2,\ldots,j_k$ is given as
\begin{equation}
    \label{Sigma_general_k_qubits}
    \Sigma_{j_1,j_2,\ldots,j_k} = \sum_i \overline{\Tr\left\{\sigma Y_i^{(j_1,j_2,\ldots,j_k)}\right\}} Y_i,
\end{equation}
where $ Y_i^{(j_1,j_2,\ldots,j_k)}$ denotes the reduced attractor $Y_i$ of qudits  $(j_1,j_2,\ldots,j_k)$. Again, this global observable $\Sigma_{j_1,j_2,\ldots,j_k}$ whose mean value is measured can vary for different choices of $k$ qudits in general. 

Third, Theorem \ref{theorem:local_observable} can be generalized beyond networks with random unitary evolution (\ref{random evolution of network}). Without going into detail, it applies without any modification to networks undergoing a discrete unital Markov evolution
\begin{equation}
    \rho(t+1) = \sum_{e \in \set E(\set G)} K_e \rho(t) K_e^{\dagger},
\end{equation}
with Kraus operators $K_e$ following the unital condition $\sum_e K_e K_e^{\dagger} = \sum_e K_e^{\dagger}K_e = I$. From that perspective, random unitary evolution (\ref{random evolution of network}) is a particular case with Kraus operators defined as $K_e = \sqrt{p_e} U_e$.

Finally, a natural and very important question is whether Theorem  \ref{theorem:local_observable} can also be utilized to design a procedure for solving the inverse problem: Assume, there is a given piece of preserved global information about the network state. Can this information be accessed locally from the chosen node $j$, and if so, how to select the corresponding local observable whose mean value provides this information? This can indeed be achieved using the following procedure: 
\begin{itemize}
    \item[1.] The preserved information is associated with a particular integral of motion $\Sigma$, which corresponds to a stationary attractor of the random unitary dynamics. Construct an orthonormal basis $\{X_{\lambda,i}\}$ of all attractors in a way that it contains this integral of motion. Let $\{\Sigma,Y_1,Y_2,\ldots,Y_m\}$ denote the subset of stationary attractors. 
    \item[2.] Find the general form of local observable $\sigma$ satisfying conditions of zero overlap (\ref{zero overlaps}) with all non-stationary attractors. This provides a parametrized set of local observables $\sigma(\alpha_1,\alpha_2,\ldots,\alpha_r)$.
    \item[3.] Calculating the reduced attractors computed for the $j$-th node and using (\ref{Sigma_general}), a general global observable $\Sigma_j=\gamma \Sigma + \sum_i \gamma_i Y_i$ is constructed for the parametrized set of local observables. Its coefficients $\gamma,\gamma_1,\ldots,\gamma_m$ linearly depend on parameters $(\alpha_1,\alpha_2,\ldots,\alpha_r)$.
    \item[4.] Now, if these parameters $(\alpha_1,\alpha_2,\ldots,\alpha_r)$ of the local observable $\sigma$ can be chosen so that all coefficients $\gamma_1,\gamma_2,\ldots,\gamma_m$ vanish, then the global information can be retrieved locally via measuring the mean value of local observable $\sigma$ with those selected parameters $(\alpha_1,\alpha_2,\ldots,\alpha_r)$. 
\end{itemize}
 
This approach will be used repeatedly in the following sections. In particular, the method is both efficient and straightforward to apply in low-dimensional attractor spaces. When multiple attractors are present, the resulting numerous contributions to a constructed global observable make it more difficult to design a desired global observable.

\section{CNOT Networks}
\label{chap:CNOTs}


The first class of networks to which we apply our theory consists of qubit networks with CNOT interactions, that is, networks in which a randomly applied two-qubit unitary operation is given by the CNOT gate. The CNOT gate is defined through its action on the computational two-qubit basis as
\begin{equation}
\label{cnot_gate}
U |ij\>= CNOT |ij\> = |i\>|i \oplus j\>,
\end{equation}
where $\oplus$ denotes addition modulo 2 and $|ij\> = |i\>|j\>$ ($i,j \in {0,1}$). Since the CNOT gate is non-symmetric, the edges of the qubit network representing qubit–qubit interactions are directed, with the tail corresponding to the control qubit and the head to the target qubit (see Fig. \ref{fig:open_network}).

There are several reasons for this choice. First, the asymptotic dynamics of such networks is well understood \cite{ruo2,Novotny2011}. Under reasonable conditions, the attractor space is low-dimensional, has no non-stationary attractors, and has a simple structure enabling the construction of integrals of motion for non-classical and non-additive global observables. Second, the model illustrates that some global information can be accessed locally by a single node, while other global observables require joint measurements across multiple nodes. Third, since the CNOT gate is a standard two-qubit unitary operation, CNOT-based networks are natural candidates for implementation on quantum computing platforms. 

\subsection{Attractors of CNOT network}
\label{sec:attractors of CNOT network}

Attractors of CNOT networks were intensively studied in \cite{Novotny2011}. It was shown that 
for strongly connected networks with $n$ qubits ($n >2$), there are two common invariant states
\begin{eqnarray}
\label{common_eigenstates_cnot}
    |\psi_0\> &=& |0,0,\ldots ,0\>, \quad |\psi_1\> = \frac{1}{\sqrt{2^n-1}}\sum_{\vec{i}\neq \vec{0}} |\vec{i}\>,
\end{eqnarray}
satisfying $U_e |\psi_i\>=|\psi_i\>$ for any two-qubit CNOT gate acting on qubits represented by the edge $e$. We denote an element of $n$-qubit computational basis as $|\vec{i}\> = |i_1, \ldots, i_n\>$, where each $i_j \in \{0,1\}$.
Using (\ref{common_eigenstates_cnot}), orthonormal basis of attractors can be constructed as
\begin{eqnarray}
\label{cnot_attractors}
    Y_1 &=& |\psi_0\>\<\psi_0|,\quad  Y_2 = |\psi_1\>\<\psi_1|, \nonumber \\
    Y_3 &=& |\psi_0\>\<\psi_1|,\quad Y_4 = |\psi_1\>\<\psi_0|, \nonumber \\
    Y_5 &=& \frac{1}{\sqrt{2^n-2}} \left(I-|\psi_0\>\<\psi_0| - |\psi_1\>\<\psi_1| \right).
\end{eqnarray}

All these attractors are permutationally invariant and associated with eigenvalue 1. Therefore, according to Theorem \ref{theorem:consensus check}, the network always reaches the strongest type of consensus (\ref{symmetric state consensus attractors}). In particular, its consensus state
\begin{equation*}
\label{cnot_general_asymptotic_state}
    \rho_{as} = \sum_{i,j = 0}^1 q_{ij} |\psi_i\>\<\psi_j| + (1-q_{00}-q_{11}) Y_5
\end{equation*}
is determined by two positive parameters $q_{ii}=\<\psi_i|\rho_{as}|\psi_i\>$ and one complex parameter $q_{01}=\overline{q_{10}}= \<\psi_0|\rho_{as}|\psi_1\>$. The three parameters describe the preserved information and uniquely determine the network’s global state after consensus has been established within the CNOT network.

\subsection{Local observables for CNOT networks}
\label{sec:local observables for CNOT netowrks}

In this section, we examine which global information about the state of the network can be accessed locally by individual nodes. Through illustrative examples, we demonstrate that while certain global properties can be inferred from local measurements on single nodes, others remain inaccessible. Following the procedure outlined at the end of Section \ref{sec:attractors of CNOT network}, we first compute all one-qubit reduced operators of the attractors in (\ref{cnot_attractors})
\begin{eqnarray}
    Y_1^{(j)} &=& |0\>\<0|, \nonumber \\
    Y_2^{(j)} &=&\frac{ 2^{n-1}-1}{2^n-1} \left(\begin{array}{cc}
        1 &  1 \\
        1 &  1+(2^{n-1}-1)^{-1}
    \end{array}\right),\nonumber \\
    Y_3^{(j)} &=& \frac{1}{\sqrt{2^n-1}}|0\>\<1|,\quad Y_4^{(j)} = \frac{1}{\sqrt{2^n-1}}|1\>\<0|,  \nonumber \\
    Y_5^{(j)} &=&\frac{2^{n-1}-1}{(2^n-1)\sqrt{2^n-2}} \left(\begin{array}{cc}
        2^n-2 & -1 \\
        -1 & 2^n
    \end{array}\right),
\end{eqnarray}
which do not depend on the choice of the qubit $j$.
Consider a general one-qubit local observable
\begin{equation}
\label{general_local_observable}
    \sigma = \left(
      \begin{array}{cc}
          a & c \\
          \overline{c} & b
      \end{array}
    \right)
\end{equation}
with real parameters $a,b$ and a complex parameter $c$.
Using (\ref{Sigma_general}), we construct for each local observable $\sigma$ the corresponding global observable $\Sigma$, given by
\begin{eqnarray}
\label{general_global_observable_cnot}
     &&\Sigma = a |\psi_0\>\<\psi_0| + \frac{c}{\sqrt{2^n-1}}|\psi_0\>\<\psi_1| + \frac{\overline{c}}{\sqrt{2^n-1}}|\psi_1\>\<\psi_0|  \nonumber \\
   &&+ \frac{1}{2^n-1}\left[(a+ 2\Re c)(2^{n-1}-1)+b2^{n-1}\right] |\psi_1\>\<\psi_1| + \nonumber \\
   && \frac{a(2^n-2)+b2^n-2\Re c}{2(2^n-1)} \left(I-|\psi_0\>\<\psi_0|-|\psi_1\>\<\psi_1|\right). \nonumber \\
\end{eqnarray}
Formula (\ref{general_global_observable_cnot}) provides all possible integrals of motion of the entire network dynamics whose mean values the individual nodes can read by measuring a corresponding local observable (\ref{general_local_observable}).

\begin{figure}[htbp]
	\centering
	\includegraphics[width=0.47\textwidth]{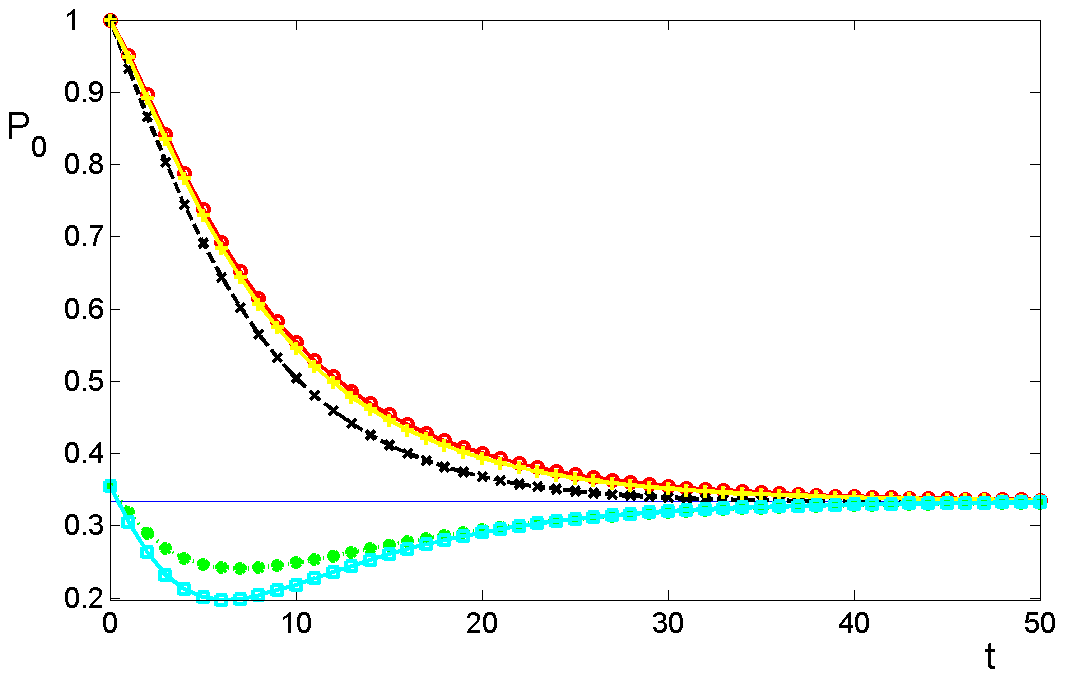}
	\caption{\centering Formation of quantum consensus in a fully connected 5-qubit CNOT network, each CNOT applied with probability $p_e=0.05$. The initial state is $|\phi>=1/\sqrt{3}|000\> \otimes (|00>+|01>+|10>)$. The blue solid line shows the preserved ground-state probability $P_0$; the other lines show the mean of observable (\ref{local observable cnot}) measured by individual nodes: 1st (solid, red, o), 2nd (dashed, black, x), 3rd (solid, yellow, +), 4th (dotted, green,*), and 5th (solid, cyan, $\square$).}
	\label{fig:cnot_nonsensus}
\end{figure}

We can now determine, for a given global observable, whether it can be accessed through local measurements performed by individual nodes. As a first example of global information, consider the probability $P_0$ that the entire qubit network is in the ground state $|0,0,\ldots,0\>$. This quantity corresponds to the coefficient $q_{00}$ in the asymptotic state (\ref{cnot_general_asymptotic_state}), and we therefore seek a local observable $\sigma$ such that
\begin{equation*}
    P_0=q_{00}=\Tr\{\rho(0)|0,0,\ldots,0\>\<0,0,\ldots,0|\} = \Tr\{\sigma \rho_{as}^{(j)}\}.
\end{equation*}
In order to obtain the global observable $\Sigma = |0,0,\ldots,0\>\<0,0,\ldots,0|$, we must set $a=1$ and require all remaining coefficients in the linear combination (\ref{general_global_observable_cnot}) to vanish, which leads to
\begin{equation*}
c=0, \quad b=\frac{1}{2^{n-1}}-1.
\end{equation*}
We thus find that there exists a unique local observable associated with the global observable $\Sigma=|\psi_0\>\<\psi_0|$,
\begin{eqnarray}
\label{local observable cnot}
\sigma=\left(
\begin{array}{cc}
1 & 0 \\
0 & \frac{1}{2^{n-1}} -1
\end{array}
\right),
\end{eqnarray}
whose measurement by individual nodes yields the probability that the entire network is in the ground state. We point out that the measurement of the mean values of (\ref{local observable cnot}) provides the true probability only if the asymptotic evolution is achieved. For some initial states this mean value can be even negative in the transient regime. This result is further corroborated by numerical simulations, as illustrated in Fig. \ref{fig:cnot_nonsensus}. We consider a CNOT network consisting of five qubits with a complete interaction graph. Each edge in the network is assigned an equal probability $p_e = 1/20$. The network is initialized in the pure state $\rho(0)=|\phi\>\<\phi|$, where $|\phi\>=1/\sqrt{3}(|00000\>+|00001\> + |00010\>)$. We find that the expectation values of the observable (\ref{local observable cnot}), measured independently by all five nodes, converge to the preserved probability $1/3$ that the network is in the ground state.

Similarly, one can demonstrate that individual nodes are able to extract partial information about the coherences between the states $|\psi_0\>$ and $|\psi_1\>$. Specifically, each node can locally access the expectation value of the global observable $\Sigma = i(|\psi_1\>\<\psi_0| - |\psi_0\>\<\psi_1|)$. This follows from the fact that such a global observable can be reconstructed when the coefficients multiplying the terms $|\psi_1\>\<\psi_0|$ and $|\psi_0\>\<\psi_1|$ are chosen proportional to $i\sqrt{2^n-1}$, while all remaining coefficients in the linear combination (\ref{general_global_observable_cnot}) are set to zero. This implies
\begin{equation}
    a=b=\Re c = 0, \quad \Im c= \sqrt{2^n-1}
\end{equation}
and we find that the corresponding local observable must be chosen proportional to the Pauli matrix $\sigma_y$, i.e. 
$$\sigma = \sqrt{2^n-1}\sigma_y = \sqrt{2^n-1} \left(\begin{array}{cc}
    0 & -i \\
    i & 0
\end{array}\right).$$ 
With this choice, we have
\begin{eqnarray}
    \Tr\{\sigma \rho_{as}^{(j)}\}&=&\Tr\{\rho(0)\Sigma\} = i \left(\<\psi_0, \rho(0) \psi_1\> - \<\psi_1, \rho(0) \psi_0\> \right) \nonumber \\ &=& 2i \Im \<\psi_0, \rho(0) \psi_1\>,
    \label{cnot_mean_value}
\end{eqnarray}
showing that individual nodes can obtain the imaginary part of the transition amplitude between states $|\psi_0\>$ and $|\psi_1\>$. In other words, individual nodes can retrieve an imaginary part of the coefficient $q_{01}$ of the global network state (\ref{cnot_general_asymptotic_state}).

In contrast, there are also pieces of global information that are inaccessible to local one-qubit nodes. In particular, the nodes cannot retrieve information about coefficient $q_{11}$ and the real part of coefficient $q_{01}$ in (\ref{cnot_general_asymptotic_state}). This means that individual nodes can not obtain the probability $P_1$ that the whole network is in the state $|\psi_1\>$ and also the real part of the transition amplitude between states $|\psi_0\>$ and $|\psi_1\>$. Indeed, for example in the case of $P_1$ we need $a=c=0$ in (\ref{general_global_observable_cnot}), implying that also $b=0$. Thus, there is no local observable (\ref{general_local_observable}) satisfying $P_1=\Tr\{\rho(0)|\psi_1\>\<\psi_1|\}=\Tr\{\rho_{as}^{(j)}\sigma\}$. Similarly, this can be shown for the real part of the transition amplitude between states $|\psi_0\>$ and $|\psi_1\>$.

\begin{figure}[htbp]
	\centering
	\includegraphics[width=0.47\textwidth]{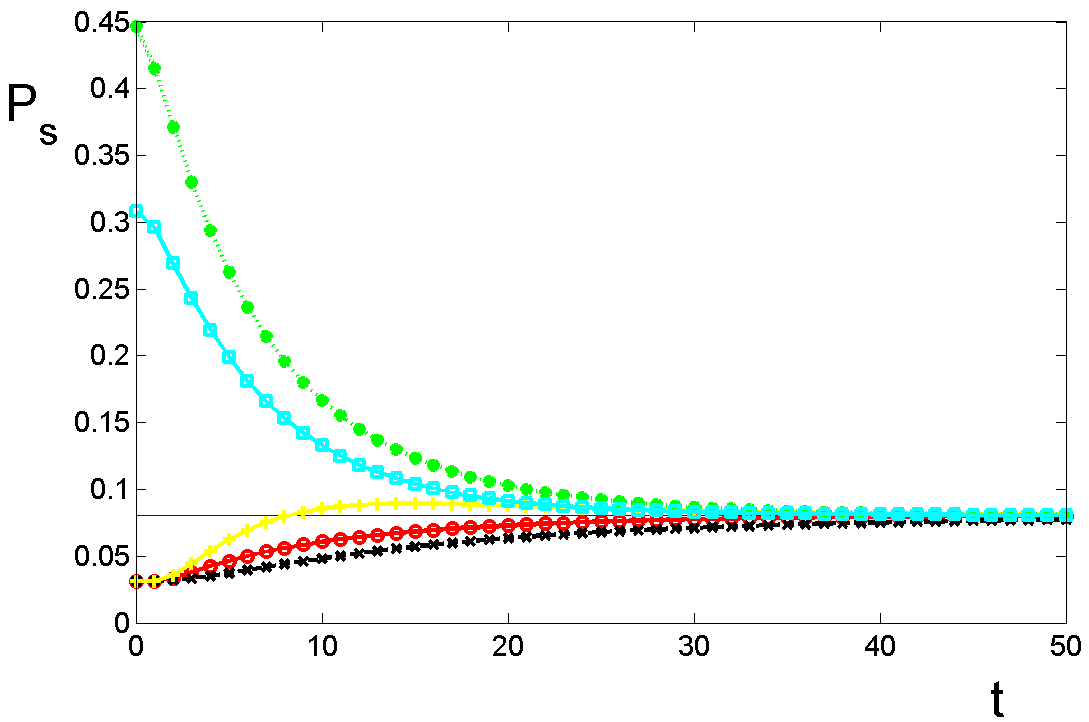}
	\caption{\centering Formation of quantum consensus in a fully connected 5-qubit CNOT network, each CNOT applied with probability $p_e=0.05$. The initial state is $|\phi>=1/\sqrt{14}|000\> \otimes (|00>+2|01>+3|10>)$. The blue solid line shows the preserved probability $P_s$ of the state (\ref{superposition_state}); the other lines show the mean of observable (\ref{local observable cnot}) measured by individual nodes: 1st (solid, red, o), 2nd (dashed, black, x), 3rd (solid, yellow, +), 4th (dotted, green,*), and 5th (solid, cyan, $\square$).}
	\label{fig:cnot_consensus_equal_superposition}
\end{figure}

Surprisingly, there is another interesting global property, which can be measured locally, provided the consensus within CNOT network is achieved. Individual nodes can get the  probability $P_s$ that the qubit network is in the equal superposition of all computational states
\begin{equation}
\label{superposition_state}
    |\psi_s\> = \left[\frac{1}{\sqrt{2}} \left(|0\>+|1\>\right)  \right]^{\otimes n} = \frac{1}{\sqrt{2^n}} \sum_{\vec{i}} |\vec{i}\>.
\end{equation}
Indeed, this state is also an invariant state of the CNOT network and hence $|\psi_s\>\<\psi_s|$ is an integral of motion. To show what corresponding local observable must be chosen, we construct different orthonormal basis of common eigenstates, in particular
\begin{eqnarray}
    \label{common_eigenstates_cnot2}
    |\tilde{\psi}_1\> &=& |\psi_s\> = \frac{1}{\sqrt{2^n}} \left(|\psi_0\> + \sqrt{2^n-1}|\psi_1\> \right), \nonumber \\
    |\tilde{\psi}_2\> &=&  \frac{1}{\sqrt{2^n}} \left(|\psi_1\> - \sqrt{2^n-1}|\psi_0\>\right).
\end{eqnarray}
We can express basis of attractors as 
\begin{eqnarray}
    \tilde{Y}_1 &=& |\tilde{\psi}_1\> \<\tilde{\psi}_1|,  \quad \tilde{Y}_2 = |\tilde{\psi}_2\> \<\tilde{\psi}_2|, \nonumber \\
     \tilde{Y}_3 &=& |\tilde{\psi}_1\> \<\tilde{\psi}_2|,  \quad \tilde{Y}_4 = |\tilde{\psi}_2\> \<\tilde{\psi}_1|, \nonumber \\
     \tilde{Y}_5 &=& \frac{1}{\sqrt{2^n-2}}\left(I - |\tilde{\psi}_1\> \<\tilde{\psi}_1| - |\tilde{\psi}_2\> \<\tilde{\psi}_2| \right). \nonumber
\end{eqnarray}
Now, calculating their one-qubit reduced operators and constructing via (\ref{Sigma_general}) the global observable $\Sigma$ for a general local observable (\ref{general_local_observable})
one can find that the only possibility to construct a global observable $\Sigma = |\psi_s\>\<\psi_s|$ is to choose local observable in the form
\begin{equation}
\label{cnot_sigma_equal_superposition}
    \sigma = \frac{1}{2^n}\left(
      \begin{array}{cc}
          1 & 2^n-1 \\
          2^n-1 & 1
      \end{array}
    \right).
\end{equation}
Numerical simulations depicted in Fig. \ref{fig:cnot_consensus_equal_superposition} for the fully connected CNOT network with 5 qubits confirm this result. All the mean values of local observable (\ref{cnot_sigma_equal_superposition}) measured by individual nodes converge to probability $P_s$. 

\begin{figure}[htbp]
	\centering
	\includegraphics[width=0.47\textwidth]{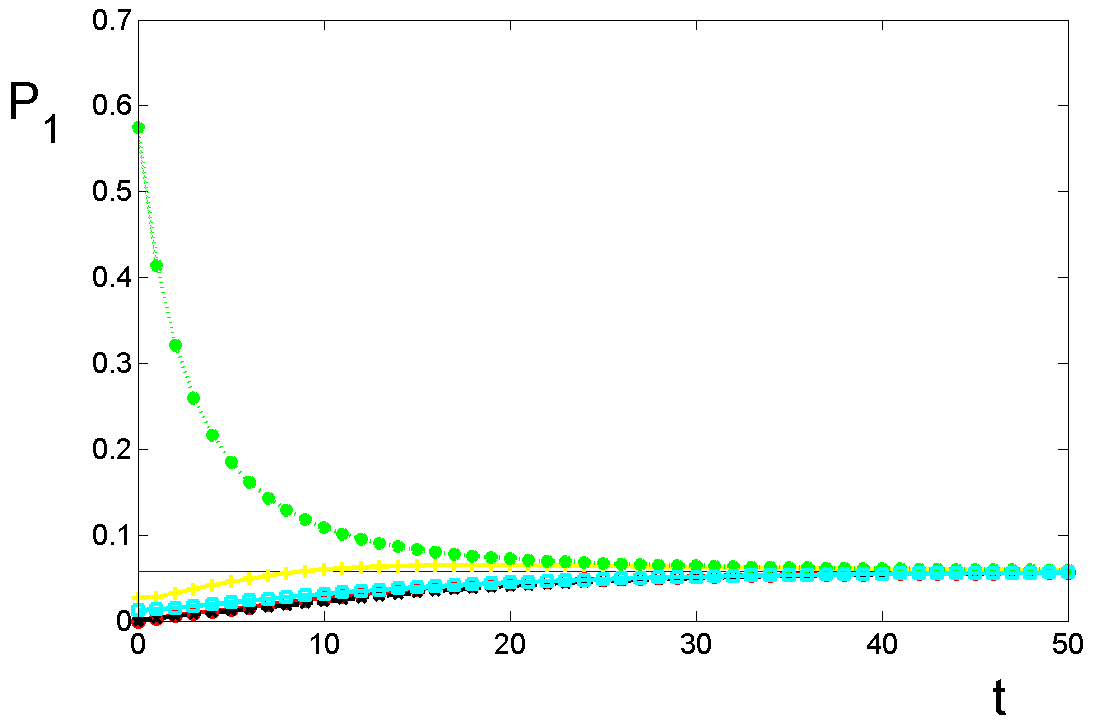}
	\caption{\centering Formation of quantum consensus in a fully connected 5-qubit CNOT network, each CNOT applied with probability $p_e=0.05$. The initial state is $|\phi>=1/\sqrt{14}|000\> \otimes (|00>+2|01>+3|10>)$. The blue solid line shows the preserved probability $P_1$ of the state $|\psi_1\>$; the other lines show the mean of two-qubit observable (\ref{general_twoqubit_local_observable}) measured by pairs of nodes: (1,2) qubits (solid, red, o), (2,3) qubits (dashed, black, x), (3,4) qubits (solid, yellow, +), (4,5)qubits (dotted, green,*), (1,5) qubits (solid, cyan, $\square$).}
	\label{fig:cnot_consensus_two_qubits}
\end{figure}

We have seen that, although a single qubit node can locally access a portion of the information preserved in the global network state, certain information remains inaccessible to an individual node. This naturally leads to the question of whether cooperation between nodes could enable retrieval of missing information. In particular, we ask whether two cooperating nodes are sufficient to determine the probability $P_1$. To address this question, we follow the same procedure as before, with the only modification being that we now consider a general two-qubit local observable
\begin{equation}
\label{general_twoqubit_local_observable}
    \sigma = \left(
      \begin{array}{cccc}
          a & e & f & g \\
          \overline{e} & b & h & k \\
          \overline{f} & \overline{h} & c & m \\
          \overline{g} & \overline{k} & \overline{m} & d
      \end{array}
    \right)
\end{equation}
with $a,b,c,d$ real and all other parameters complex.
Repeating the same steps while using (\ref{Sigma_general}) for the two-qubit reduced attractors, we derive the general form of the global observable $\Sigma$. It takes the same shape as (\ref{general_global_observable_cnot}) but with coefficients defined in terms of matrix elements (\ref{general_twoqubit_local_observable}). In order to obtain a global observable capable of measuring probability $P_1$ we find that while the coefficient corresponding to term $|\psi_1\>\<\psi_1|$ must be equal to one, all the other must be zero. A straightforward algebra reveals that it is accomplished if and only if 
\begin{eqnarray}
    a&=&0, \quad e+f+g=0, \nonumber \\
    b+c+d&=& \frac{4}{2^n}, \quad \Re(h+k+m) = 4 \frac{2^{n-1}-1}{2^n}.
\end{eqnarray}
We find that two nodes can indeed retrieve probability $P_1$ if they are allowed to measure together. It is interesting that in this case the solution is not unique. Fig. \ref{fig:cnot_consensus_two_qubits} shows numerical results for the fully connected 5-qubit network. The two-qubit observable (\ref{general_twoqubit_local_observable}) is chosen with parameters $e=f=g=0$, $b=c=d=1/24$ and $h=k=m=15/24$. It confirms that all chosen two-qubit joint measurements of its mean value converge to probability $P_1$.


\section{Quantum average consensus}
\label{sec:total number of excitation}

In this section, we demonstrate that the general framework introduced in Section \ref{sec:Random unitary evolution towards consensus} can also be used in a reversed setting, namely for designing network interactions that realize a prescribed redistribution of information across the network. We illustrate this idea through the concept of quantum average consensus (QAC), the quantum analog of the classical average consensus protocol. In the classical case, nodes initialized with individual values repeatedly exchange information with their neighbors and update their states until they converge to a common value equal to the average of the initial values. The quantum counterpart aims to redistribute the same type of information. Any initial quantum state of the network encodes a specific distribution of excitations among nodes. The QAC protocol is therefore designed to preserve the average excitation number while steering the network toward a consensus state in which this information may become locally accessible at each node. We note that measuring the average number of excitations per node is equivalent to measuring the total number of excitations in the network when the network size is known. Consequently, redistributing either quantity constitutes the same task.

At first glance, this task could also be approached using a classical strategy: each node would measure the mean excitation number, after which a classical average consensus protocol could be applied. However, this approach is fundamentally different, as it requires local measurements of the excitation number -- quantities that may not be directly accessible -- and subsequently relies on a classical network algorithm. In contrast, the quantum protocol operates directly at the level of quantum states. Once consensus is reached, individual nodes possess local states from which they may read out the average excitation number, and in some cases may also access additional information encoded in the final state of the network (see Fig. \ref{fig:classical_versus_quantum_average_consensus}).

The central questions addressed in this section are which types of interactions can achieve this task independently of the network size and geometry, and what additional information can be accessed locally by one or more nodes once QAC has been reached. 
Two properties must be satisfied in networks with such interactions:
\begin{itemize}
    \item[1.] The total number of excitations in the network must be an integral of motion.
    \item[2.] The dynamics drives the network into an asymptotic regime where this information is locally accessible to each node independently at sufficiently long times.
\end{itemize}
We intend to investigate these questions for qubit networks. Such mechanism must work for large networks as well as for the smallest one consisting of only two nodes.

\subsection{Consensus for a two-qubit network}
\label{sec:consensus_two-qubit_network}
Let us therefore first consider a network of two qubit nodes with two edges $(1,2)$ and $(2,1)$. The corresponding RUO governing its evolution is
\begin{equation}
\label{two_qubit_unitary_average}
    \rho(t+1)= p_{12}U_{12}\rho(t)U_{12}^{\dagger}+p_{21}U_{21}\rho(t)U_{21}^{\dagger}
\end{equation}
with probabilities satisfying $p_{12}+p_{21}=1$.

In the first step, we are looking for two-qubit unitary gates $U_{12}$ (and analogously for $U_{21}$), which preserve the mean value of the number of excitations. Hence, the two-qubit operator of the number of excitations
\begin{equation}
    \label{operator_number_excitation}
    N_2=|01\>\<01|+|10\>\<10|+2|11\>\<11|
\end{equation}
must be an integral of motion for such two-qubit dynamics. For a two-qubit network, it means that unitaries $U_{12}$ and $U_{21}=\Pi_{12} U_{12} \Pi_{12}$ (where $\Pi_{12}$ is the exchange operator changing the order of the first and second qubit)  must satisfy commutation relations
\begin{equation}
    [U_{12},N_2]=[U_{21},N_2]=0.
\end{equation}
Therefore, $U_{12}$ and $U_{21}$ must have a matrix block structure
\begin{equation}
\label{def_two_qubit_unitary}
    U_{12}(U_{21}) = \left(\begin{array}{ccc}
        V_1 & 0 & 0  \\
        0 & V_2(\tilde V_2) & 0 \\
        0 & 0 & V_3  
    \end{array} \right)
\end{equation}
with one-dimensional blocks $V_1=\alpha_0$ and $V_3=\alpha_3$ and 
\begin{equation}
\label{def_two_qubit_unitary_21}
V_1 = \left(\begin{array}{cc}
        u_{11} & u_{12}  \\
        u_{21} & u_{22} \\
    \end{array} \right), \quad V_2 = \left(\begin{array}{cc}
        u_{22} & u_{21}  \\
        u_{12} & u_{11} \\ 
    \end{array} \right).
\end{equation}
Note that we consider only non-symmetric interactions, i.e. $V_2 \neq \tilde{V}_2$, because otherwise the whole evolution of two-qubits is unitary bringing the state of the network arbitrarily close to the initial state and thus prohibiting quantum consensus.
These are all 2-qubit unitary interactions that keep the mean number of excitations constant in an arbitrarily large network. 

\subsubsection{Attractor analysis for two-qubit network}
\label{sec:two-qubit mechanisms}
To choose interactions which satisfy also the second property we must analyze attractor structure of dynamics generated by these interactions. Attractor equation (\ref{eq_attractors_general}) for $U_{12}$ and a general 2-qubit attractor $X$ with block structure
\begin{equation}
    X= \left(\begin{array}{ccc}
        x_{11} & x_{12} & x_{13}\\
        x_{21} &x_{22} & x_{23} \\
        x_{31} & x_{32} & x_{33}
    \end{array} \right)
\end{equation}
divides into independent equations 
\begin{equation}
\label{eq_block_attractor_equation}
   V_i x_{ij} = \lambda x_{ij} V_j. 
\end{equation}
Thus, each nonzero block $x_{ij}$ following (\ref{eq_block_attractor_equation}) defines a corresponding attractor, where all other blocks of the attractor are taken as zero. Since $V_1$ and $V_3$ are one-dimensional blocks, there is always an attractor corresponding to the one-dimensional block $x_{13}=1$ (resp. $x_{31}=1$) associated with eigenvalue $\lambda= \alpha_0\bar{\alpha}_3$ (resp. $\lambda= \alpha_3\bar{\alpha}_0$). The off-diagonal ($2 \times 1$)-dimensional block $x_{21}$ must satisfy equation $V_2 x_{21} = \lambda \alpha_0 x_{21}$ and simultaneously the same equation for $\tilde{V}_2$. Thus $x_{21}$ is a common eigenvector of both unitaries $V_2$ and $\tilde{V}_2$ for the same eigenvalue. Since this inevitably means the contradiction $V_2 = \tilde{V}_2$, we find $x_{21}=0$. Analogously, one can show that off-diagonal blocks $X_{21},X_{23}$ and $X_{32}$ must be zero, too. 

Apparently, diagonal one-dimensional blocks $x_{11}$ and $x_{33}$ define two attractors each corresponding to the eigenvalue $\lambda =1$. 

The last middle diagonal block $x_{22}$ is determined by equation $V_2 x_{22} = \lambda x_{22} V_2$ and the analogous equation for $\tilde{V}_2$ with the only difference that the role of the states $|0\>$ and $|1\>$ is interchanged. Therefore, we can construct this block using eigenvectors of unitaries $V_2$ and $\tilde{V}_2$. Indeed, both equations are eigenvalue equations of the form $VXV^{\dagger}=\lambda X$. Assume general orthonormal base eigenvectors of $V_2$ written as
\begin{eqnarray}
   |e_1\> &=& \cos(\phi) |01\> + \sin(\phi) e^{i\delta} |10\> \\
   |e_2\> &=& \sin(\phi) |01\> - \cos(\phi) e^{i\delta} |10\>
\end{eqnarray}
associated with eigenvalue $\alpha_1$ and $\alpha_2$ respectively.
Then corresponding orthonormal base eigenvectors of $\tilde{V}_2$ read 
\begin{eqnarray}
   |\tilde{e}_1\> &=& \sin(\phi)|01\>+\cos(\phi)e^{-i\delta}  |10\> \\
|\tilde{e}_2\> &=& \cos(\phi) |01\> - \sin(\phi) e^{-i\delta} |10\> 
\end{eqnarray}
associated with eigenvalue $\alpha_1$ and $\alpha_2$ respectively.
Consequently, orthonormal base eigenmatrices of the linear map $\map V_2(X) = V_2XV_2^{\dagger}$ with respect to the Hilbert-Schmidt norm are $|e_1\>\<e_1|$ and $|e_2\>\<e_2|$(associated with eigenvalue 1) and $|e_1\>\<e_2|$ and $|e_2\>\<e_1|$ (associated with eigenvalue $\alpha_1\bar{\alpha}_2$ and $\alpha_2\bar{\alpha}_1$ respectively) and analogously for the linear map $\tilde{\map V}_2(X) = \tilde{V}_2X\tilde{V}_2^{\dagger}$. Any attractor associated with the nonzero block $x_{22}$ must solve equation $VXV^{\dagger}=\lambda X$ simultaneously for both unitaries $V_2$ and $\tilde{V}_2$ and common eigenvalue $\lambda$. Apparently, since both $V_2$ and $\tilde{V}_2$ are unitaries, the operator $|e_1\>\<e_1|+|e_2\>\<e_2|= |\tilde{e}_1\>\<\tilde{e}_1|+|\tilde{e}_2\>\<\tilde{e}_2|$ is always an attractor corresponding to eigenvalue $1$. Whether there are more attractors associated with the nonzero block $x_{22}$ depends on mutual symmetries of eigenstates and eigenvalues of $V_2$ and $\tilde{V}_2$.

Let us first treat two possible symmetries of their eigenstates. The first one is for $\phi=\pi/4$ and $e^{2i\delta}=1$, for which we get $|e_1\>=|\tilde{e}_1\>$ and $|e_2\>=|\tilde{e}_2\>$. This implies that $U_{12}$ is symmetric and the attractor space is 16-dimensional containing the whole operator Hilbert space $\map B(\hil H)$. The second symmetry occurs when $e^{2i\delta}=-1$, i.e. $\cos(\delta)=0$, for which we find $|e_1\>=|\tilde{e}_2\>$ and $|e_2\>=|\tilde{e}_1\>$. From that follows that $|e_1\>\<e_1|=|\tilde{e}_2\>\<\tilde{e}_2|$ and $|e_2\>\<e_2|=|\tilde{e}_1\>\<\tilde{e}_1|$ are eigenmatrices all corresponding to eigenvalue $\lambda=1$. They constitute two linearly independent attractors each corresponding to eigenvalue $\lambda=1$. However, note that this is just one additional attractor, since $|e_1\>\<e_1|+|e_2\>\<e_2|$ is always attractor. Similarly we have also $|e_1\>\<e_2|=|\tilde{e}_2\>\<\tilde{e}_1|$ and $|e_2\>\<e_1|=|\tilde{e}_1\>\<\tilde{e}_2|$, but in general they correspond to different eigenvalues $\alpha_1 \bar{\alpha}_2$ and $\alpha_2 \bar{\alpha}_1$, thus in general they do not correspond to a common solution for the same eigenvalue $\lambda$ and can not be an attractor.

Symmetries of spectra of $V_2$ and $\tilde{V}_2$ leading to existence of more attractors are also two. They both correspond to an additional degeneration in spectra of maps $\map V$ and $\tilde{\map V}$, when $\alpha_1\bar{\alpha}_2$ and $\alpha_2\bar{\alpha}_1$ are equal. The first case occurs when $\alpha_1=\alpha_2$ which results in $V_2 = \alpha_1 I = \tilde{V}_2$, i.e. symmetric unitary $U_{12}$, with attractor space equal to the full operator space $B(\hil H)$. The second case $\alpha_1=-\alpha_2$ is more involved. Both eigenmatrices $|e_1\>\<e_2|$ and $|e_2\>\<e_1|$ as well as $|\tilde{e}_1\>\<\tilde{e}_2|$ and $|\tilde{e}_2\>\<\tilde{e}_1|$ correspond to the same eigenvalue $\lambda=-1$ of the map $\map V_2$ and $\tilde{\map V}_2$ respectively. We can construct an attractor from them if spans of these eigenmatrices have overlap, i.e. if there are nonzero complex coefficients $a_1$, $b_1, a_2$, and $\b_2$ such that
$$
a_1 |e_1\>\<e_2| + b_1 |e_2\>\<e_1| = a_2 |\tilde{e}_1\>\<\tilde{e}_2| + b_2 |\tilde{e}_2\>\<\tilde{e}_1|.
$$
This gives us three conditions
\begin{eqnarray}
    a_1 +b_1 &=& a_2 + b_2, \nonumber \\
   a_2 \sin^2(\phi) e^{2i\delta}- b_2 \cos^2(\phi) e^{2i\delta}  &=& a_1 \cos^2(\phi) - b_1 \sin^2(\phi) \nonumber \\
     a_1 \sin^2(\phi) e^{2i\delta}- b_1 \cos^2(\phi) e^{2i\delta} &=& a_2 \cos^2(\phi)- b_2 \sin^2(\phi). \nonumber
\end{eqnarray}
Subtracting the last two equations and applying the first one we obtain
\begin{equation}
    (a_1-a_2)\left(1+e^{2i\delta}\right)=0.
\end{equation}
Let us treat the case $e^{2i\delta}\neq -1$ first, for which we have $a_1=a_2$ and $b_1=b_2$ implying
\begin{equation}
    a_1 (\sin^2(\phi)e^{i\delta}-\cos^2(\phi)e^{-i\delta})= b_1 (\cos^2(\phi)e^{i\delta}-\sin^2(\phi)e^{-i\delta}). \nonumber
\end{equation}
Choosing $a_1=\cos^2(\phi)e^{i\delta}-\sin^2(\phi)e^{-i\delta}$ and $b_1 = \sin^2(\phi)e^{i\delta}-\cos^2(\phi)e^{-i\delta}$ we receive a new normalized attractor corresponding to eigenvalue $\lambda=-1$
\begin{equation}
     \frac{1}{K} (a_1 |e_1\>\<e_2| + b_1 |e_2\>\<e_1|) =  \frac{1}{\sqrt{2}} (|01\>\<10| - |10\>\<01|). \nonumber
\end{equation}
In a special case $\alpha_1=-\alpha_2$ and $e^{2i\delta}=-1$ when symmetries of eigenstates and eigenvalues occur simultaneously, we get altogether 4 attractors associated with the nonzero block $x_2$: $|e_1\>\<e_1|, |e_2\>\<e_2|$ corresponding to eigenvalue $\lambda=1$ and $|e_1\>\<e_2|, |e_2\>\<e_1|$ corresponding to eigenvalue $\lambda=-1$.

Let us summarize our analysis of all possible two-qubit attractor spaces for different parameter values of unitaries $U_{12}$ and $\tilde{U}_{21}$ with $V_2$ parametrized as 
\begin{equation}
    \label{V2 parametrized}
    V_2 = \left( 
    \begin{array}{cc}
         \alpha_1-\sin^2(\phi)(\alpha_1-\alpha_2) & \frac{1}{2}\sin(2\phi)e^{-i\delta}(\alpha_1-\alpha_2)  \\
         \frac{1}{2}\sin(2\phi)e^{i\delta}(\alpha_1-\alpha_2) &  \alpha_1-\cos^2(\phi)(\alpha_1-\alpha_2)
    \end{array}
    \right). \nonumber
\end{equation}
Its eigenvectors $|e_0\> = |00\>,|e_1\>,|e_2\>,|e_3\>=|11\>$ correspond to eigenvalues $\alpha_0,\alpha_1,\alpha_2$ and $\alpha_3$ respectively.

There are 6 parameter regions of $U_{12}$, each with a corresponding orthonormal basis of attractors enlisted in Table \ref{table:attractors_2_qubits}.
\begin{table*}[tbhp]
\centering
\caption{Analysis of attractor spaces for two-qubit network under RUO of the form (\ref{def_two_qubit_unitary}) with $V_2$ parametrized as (\ref{V2 parametrized}).}
\label{table:attractors_2_qubits}
 \begin{tabular}{||c c c c c c||} 
 \hline
 Case & Parameter region & Attractor dimension &  ON attractor basis &  Eigenvalues & Consensus\\
 \hline\hline
 1. & $\alpha_1=\alpha_2$ & 16 & $|e_i\>\<e_j|$, $i,j \in \{0,1,2,3\}$   & $\alpha_i\bar{\alpha}_j$  & No \\ 
 \hline
 2. & $\phi=\pi/4$,$\cos(\delta)=\pm 1$ & 16 & $|e_i\>\<e_j|$, $i,j \in \{0,1,2,3\}$ & $\alpha_i\bar{\alpha}_j$  & No\\
 \hline
 \multirow{3}{*}{3.} & \multirow{3}{*}{$\cos(\phi)=0$} & \multirow{3}{*}{6} & $|e_i\>\<e_i|$, $i \in \{0,1,2,3\}$ & 1 & \\
 & & & $|e_0\>\<e_3|$ & $\alpha_0\bar{\alpha}_3$ & No\\
 & & & $|e_3\>\<e_0|$ & $\alpha_3\bar{\alpha}_0$ &  \\
 \hline
\multirow{4}{*}{4.} & \multirow{4}{*}{$\alpha_1=-\alpha_2$} & \multirow{4}{*}{6} & $|e_0\>\<e_0|$, $|e_3\>\<e_3|$, $\frac{1}{\sqrt{2}}\left(|e_1\>\<e_1|+|e_2\>\<e_2| \right)$ & 1 &\\
 & & & $\frac{1}{\sqrt{2}}\left(|01\>\<10|-|10\>\<01| \right)$ & -1 & \\
 & & & $|e_0\>\<e_3|$ & $\alpha_0\bar{\alpha}_3$ & Yes\\
 & & & $|e_3\>\<e_0|$ & $\alpha_3\bar{\alpha}_0$ & \\
 \hline
\multirow{4}{*}{5.} & \multirow{4}{*}{$\alpha_1=-\alpha_2,\cos(\phi)=0$} & \multirow{4}{*}{8} & $|e_i\>\<e_i|$, $i \in \{0,1,2,3\}$ & 1 & \\
 & & & $|e_1\>\<e_2|,|e_2\>\<e_1|$ & -1 & \\
  & & & $|e_0\>\<e_3|$ & $\alpha_0\bar{\alpha}_3$ & No\\
 & & & $|e_3\>\<e_0|$ & $\alpha_3\bar{\alpha}_0$ & \\
 \hline
\multirow{3}{*}{6.} & \multirow{3}{*}{other cases} & \multirow{3}{*}{5}  & $|e_0\>\<e_0|, |e_3\>\<e_3|, \frac{1}{\sqrt{2}}\left(|e_1\>\<e_1|+|e_2\>\<e_2|\right)$ & 1 & \\
  & & &$|e_0\>\<e_3|$ & $\alpha_0\bar{\alpha}_3$ & Yes\\
 & & & $|e_3\>\<e_0|$ & $\alpha_3\bar{\alpha}_0$ & \\
 \hline
\end{tabular}
\end{table*}

\subsubsection{Quantum consensus for networks with $n=2$}
\label{sec:consensus for 2-qubits}

Using the attractor analysis of (\ref{two_qubit_unitary_average}) together with (\ref{def_two_qubit_unitary}) and (\ref{def_two_qubit_unitary_21}), we determine which mechanisms lead to consensus, i.e., an asymptotic network state encoding the mean excitation number locally. Clearly, no consensus is achieved when the attractor space is $16$-dimensional (cases 1 and 2), as the unperturbed unitary dynamics quasi-periodically returns the global state arbitrarily close to its initial configuration. Consequently, the local reduced states contain no global information about the mean excitation number.

To analyze the remaining cases, we employ the theory developed in Section~\ref{chap:Quantum Consensus Characterized by Attractors}. It follows that the one-qubit reductions of attractors corresponding to eigenvalues different from one are all equal to zero. Consequently, the first assumption of Theorem~\ref{theorem:local_observable} is satisfied.

Considering a general one-qubit observable~(\ref{general_local_observable}), we compute the associated global observable (integral of motion) given by~(\ref{Sigma_general}). For the case of four attractors associated with eigenvalue~1 ($\cos(\phi)=0$), we obtain $\Sigma=\diag(a,a,b,b)$. Since there is no choice of parameters of $\sigma$ for which $\Sigma=N_2$, the unitary interactions corresponding to the 3rd and 5th cases in Table~\ref{table:attractors_2_qubits} do not drive the network to the qubit average consensus.

In contrast, when three attractors are associated with eigenvalue~1, the corresponding global observable takes the form $\Sigma_1=\Sigma_2=\diag(a,(a+b)/2,(a+b)/2,b)$. Clearly, choosing parameters $a=0$ and $b=2$ defines a local observable whose measured mean value directly yields the total number of excitations in the network and, consequently, the average number of excitations per node. Note that the local observable is not uniquely defined, since $c\in\Cmplx$.

This also allows us to address whether individual nodes can access additional information. The answer is no: choosing $a=b=1$ and $c\in\Cmplx$ leads to the measurement of the observable $\Sigma=\diag(1,1,1,1)$, whose mean value is simply $\Tr(\rho(0))=1$. Therefore, measuring the mean value of a general $\sigma$ yields only a linear combination of these two mean values, and no further information can be extracted by local nodes. The case in which multiple qubits perform a joint measurement is discussed in Section \ref{Information gained by collective measurements of more agents of the network}.

For the minimal qubit network, we find that the desired consensus can be achieved only using two-qubit unitary gates of the form (\ref{def_two_qubit_unitary}), with parameters corresponding to the 4th and 6th cases in Table \ref{table:attractors_2_qubits}. Furthermore, it is straightforward to verify that while the 4th case drives the two-qubit network to the RSC but not to the SSC, the 6th case leads the network to the SSC for any initial state.

\subsection{Consensus for qubit networks of an arbitrary size}
\label{sec:consensus_arbitrary_size}

In this part, we focus on mechanisms capable of establishing average consensus in networks of arbitrary finite size, including $n=2$.
Since we have already found consensus mechanisms for $n=2$, it is sufficient to check which of them also work for $n\geq 2$. In both cases, in case 4 and 6 in Table \ref{table:attractors_2_qubits}, the attractor structure is always the same for $n>2$. The dimension of the attractor space is $n+3$. While two non-stationary attractors $|00 \ldots 0\>\<11 \ldots 1|$ and $|11 \ldots 1\>\<00 \ldots 0|$ do not contribute to one-qubit reduced states, the rest $n+1$ stationary attractors do. Elements of their orthonormal basis are proportional to projectors on subspaces with fixed number of excitations. In order to write down their explicit forms, let us define the number of excitations for each of the computational states $|i_1,i_2,\ldots,i_n\>$ as 
\begin{equation}
\label{def_scitani_excitaci}
s(\vec{i}) = \sum_{j=1}^n i_j. 
\end{equation}
The subset of computational basis with states of exactly $k$ excitations is $\set S_k= \{|\vec{i}\> | s(\vec{i})=k\}$. Then orthonormal basis of atractors associated with eigenvalue 1 is given by properly normalized operators
\begin{equation}
    \label{def_projections}
    A_k=  \frac{1}{\sqrt{\binom{n}{k}}}P_k= \frac{1}{\sqrt{\binom{n}{k}}}\sum_{|v\> \in \set S_k} |v\>\<v|, \quad k \in \{0,1,\ldots, n\}.
\end{equation}
$P_k$ denotes the orthogonal projector on the span of computational states with a fixed number of excitations $k$. 

This has two consequences. First, since all attractors are permutationally invariant, according to Theorem \ref{theorem:consensus check} the quantum network always reaches SSC. Therefore, the state of any subsystem of a given size is identical and all subsystems of the same size contain the same information about the network. Second, the projectors $P_k$ are integrals of motion for the overall quantum network dynamics in the Heisenberg picture (\ref{heisenberg_picture_ruo}). Their expectation values define the probability distribution
\begin{equation}
\label{probability_distribution_of_k_particles}
q_k=\Tr\{P_k \rho(0)\}= \Tr\{P_k \rho_{as}\},
\end{equation}
which gives the probability that the network contains exactly $k$ excitations, and these quantities remain invariant throughout the evolution.

Consequently, we arrive at the notable conclusion that this probability distribution — apart from coherences between the ground state and the fully excited state of the network — constitutes the only information retained in the consensus state of the network. We emphasize that this is information determining the overall state of the quantum network but not necessarily accessible for individual nodes.

Let us now verify whether the network (with $n \geq 2$) reaches the desired average consensus state, namely, whether each node can access the average number of excitations in the network. The single-qubit reduced operators of $A_k$ take the following form in the computational basis
\begin{equation}
    A_k^{(1)} = \Tr_{n-1}{A_k}=\frac{\sqrt{\binom{n}{k}}}{n} \left(\begin{array}{cc}
       n-k  & 0 \\
        0 & k
    \end{array}\right),
\end{equation}
where $\Tr_{n-1}$ denotes trace over $n-1$ qubits.

According to (\ref{Sigma_general}), any qubit node measuring a general local observable (\ref{general_local_observable}) will measure the mean value of the global observable
\begin{equation}
    \label{global+observable_average_bigger_2}
    \Sigma= a \sum_{k=0}^n \frac{n-k}{n}P_k + b \sum_{k=0}^n \frac{k}{n}P_k.
\end{equation}
One can see that the choice $a=0,b=n$, i.e. measurement of observable
\begin{equation}
    \sigma = \left(\begin{array}{cc}
       0  & c \\
        \bar{c} & n
    \end{array}\right),
\end{equation}
results in measurement of global observable 
\begin{equation}
    \Sigma_T= \sum_{k=0}^n k P_k.
\end{equation}
Its mean value is the expected value of the total number of excitations in the network
\begin{equation}
    \Tr\{\rho \Sigma_T\} = \sum_{k=0}^n k q_k = \<N_n\>.
\end{equation}
Thus, we find that all these mechanisms given by the 4th and the 6th case in Table \ref{table:attractors_2_qubits} bring the network of an arbitrary size to the consensus where all nodes can independently measure the mean value $\<N_n\>$ of the total number of excitations in the network. Note that in order to measure the average number of excitations in the network per node, each node must measure the mean value of observable 
\begin{equation}
    \sigma = \left(\begin{array}{cc}
       0  & c \\
        \bar{c} & 1
    \end{array}\right).
\end{equation}

Let us consider the same question as in the previous case $n=2$, namely whether individual nodes can extract additional information. The answer is negative. Setting $a=1,b=1$ for the local observable (\ref{general_local_observable}) corresponds to measuring the expectation value of the observable $\tilde{\Sigma}= \sum_k P_k=I$, which does not provide additional information. By linearity, the general global observable (\ref{global+observable_average_bigger_2}) can be expressed as
\begin{equation}
\Sigma = \frac{b-a}{n}\Sigma_T + a \tilde{\Sigma}.
\end{equation}
We can conclude that its expectation value contains information only about the total number of excitations in the network.

\subsection{Collective measurement of two or more nodes}
\label{Information gained by collective measurements of more agents of the network}

Unlike the classical average consensus protocol, its quantum counterpart operates at the level of quantum states. As a result, the achieved consensus state retains more information than is strictly necessary for the original task. In this section, we investigate whether nodes can leverage this preserved information through cooperation. In particular, if nodes are permitted to carry out joint two- or multi-qubit measurements, what additional information can they extract? To answer this question, we make use of Theorem \ref{theorem:local_observable} and its generalization given by (\ref{Sigma_general_k_qubits}). To identify what global information is retrieved from a joint $m$-qubit measurement of the expectation value of the local observable $\sigma$, we first determine the $m$-qubit reduced operators of the attractors (\ref{def_projections}). All these operators are diagonal, with the diagonal entry associated with the state $|\vec{i}\rangle = |i_1,i_2,\ldots,i_m\rangle$ given by 
\begin{equation}
\label{mqubit reduced projections}
\left(A^{(m)}_k\right)_{\vec{i}}= \binom{n}{k}^{-1/2} \binom{n-m}{k-s(\vec{i})},
\end{equation}
It is convenient to represent a general local $m$-qubit observable as
\begin{equation}
\label{mqubit_general_sigma_average}
\sigma = \sum_{|\vec{i}\> \in \hil H^m} \sigma_{\vec{i}} |\vec{i}\>\<\vec{i}| + \sigma_{rest},
\end{equation}
where $\sigma_{\vec{i}}$ denotes the diagonal entry of the observable $\sigma$ and $\sigma_{rest}$ its off-diagonal part. In this notation, the corresponding general global observable (\ref{Sigma_general_k_qubits}) reads
\begin{eqnarray}
\Sigma = \sum_{k=0}^{n} A_k \overline{\Tr\{A_k^{(m)}\sigma\}}= \sum_{k=0}^{n} P_k \sum_{|\vec{i}\> \in \hil H^m} \sigma_{\vec{i}} \frac{\binom{n-m}{k-s(\vec{i})}}{\binom{n}{k}}. \nonumber
\end{eqnarray}
Because the binomial coefficients depend solely on the number of excitations, it follows that
\begin{eqnarray}
\Sigma = \frac{(n-m)!}{n!}\sum_{k=0}^{n} P_k \sum_{j=0}^m T_j \frac{k!(n-k)!}{(k-j)!(n-m-k+j)!}, \nonumber
\end{eqnarray}
where $T_j=\sum_{\vec{i}:s(\vec{i})=j} \sigma_{\vec{i}}$ sums up all coefficients $\sigma_{\vec{i}}$ corresponding to states $|\vec{i}\>$ with fixed number of excitations $j$. The second summation can be further rewritten as a polynomial in the variable $k$ of order $m$
\begin{eqnarray}
\label{Sigma_m_moment}
\Sigma = \frac{(n-m)!}{n!}\sum_{k=0}^{n} P_k \sum_{l=0}^m c_l k^l = \frac{(n-m)!}{n!}\sum_{l=0}^m c_l\sum_{k=0}^{n} k^lP_k.
\end{eqnarray}

This shows that the general global observable, whose expectation value is obtained from a joint measurement of the mean value of an $m$-qubit observable, can be expressed as a linear combination of observables corresponding to the first $m$ moments of the probability distribution $q_k$. Thus, a collective measurement on any set of $m$ nodes provides access to these moments, with each additional node enabling the retrieval of the next higher-order moment, as illustrated in Fig. \ref{fig:classical_versus_quantum_average_consensus}.

In the following, we show how to explicitly construct an $m$-qubit observable whose expectation value yields the $m$-th moment of the distribution $q_k$. One option is to use (\ref{Sigma_m_moment}), where all coefficients $c_l$ are set to zero, except for $c_m=n!/(n-m)!$. This results in $m+1$ linear equations involving the coefficients $T_j$ (with $j\in {0,1,m}$), which specify the admissible values of the diagonal parameters $\sigma_{\vec{i}}$.

We also present an alternative approach that provides an iterative construction of such an observable. Since all $m$-qubit reduced operators (\ref{mqubit reduced projections}) of attractors (\ref{def_projections}) are diagonal, the diagonal element of the $m$-qubit reduced state of asymptotic state (\ref{eq_asymptotic_evolution_general}) can be written as
\begin{equation}
\label{mqubit reduced state}
\left(\rho_{as}^{(m)}\right)_j = \sum_{k=0}^n q_k \binom{n}{k}^{-1} \binom{n-m}{k-j},
\end{equation}
where the index $j$ corresponds to the number of excitations in the associated computational basis state $|i_1,i_2,\ldots,i_m\>$. If we select $j=m$, i.e., focus on the diagonal matrix element of $\rho_{as}^{(m)}$ corresponding to the state vector $|1,1,\ldots,1\>$, we obtain
\begin{eqnarray}
\label{mqubit reduced state m-element}
\left(\rho_{as}^{(m)}\right)_m &=& \frac{(n-m)!}{n!} \sum_{k=0}^n q_k k(k-1) \ldots (k-m+1) \nonumber \\
&=& \frac{(n-m)!}{n!} \sum_{k=0}^n q_k \left(k^m + \sum_{j=1}^{m-1} b_j k^j \right) \nonumber \\ &=& \frac{(n-m)!}{n!} \left( \<k^m\> + \sum_{j=1}^{m-1} b_j \<k^j\> \right),
\end{eqnarray}
where $b_j$'s denote expansion coefficients of the polynomial $k(k-1)\ldots(k-m+1)$. This matrix element is therefore determined by a linear combination of the $j$-th moments of the probability distribution $q_k$ with $j \in {1,2,\ldots, m}$. Other matrix elements are also determined by these moments. Consequently, the $m$-qubit reduced state is fully specified by, and itself determines, the first $m$ moments of the distribution $q_k$.

Let $O_k$ denote the $k$-qubit observable whose expectation value gives the $k$-th moment of $q_k$. The explicit form of this observable has already been derived for the one-qubit subsystem. Observables corresponding to higher-order moments can be constructed using the following iterative procedure
\begin{equation}
\label{induction recipe}
O_m = \frac{n!}{(n-m)!} |11\ldots 1\>\<11\ldots 1| - \sum_{j=1}^{m-1} b_j O_j\otimes I_{m-j},
\end{equation}
where $I_j$ denotes the identity operator acting on the Hilbert space of $j$ qubits. Indeed, it can be checked by a straightforward calculation 
\begin{eqnarray}
    && \<O_m\>_{\rho_{as}^{(m)}} = \Tr\left\{O_m \rho_{as}^{(m)}\right\} = \nonumber \\
   && \<k^m\> + \sum_{j=1}^{m-1}  b_j \<k^j\> - \sum_{j=1}^{m-1}  b_j \Tr\left\{\rho_{as}^{(m)} O_j\otimes I_{m-j}\right\} = \nonumber \\
    && \<k^m\> + \sum_{j=1}^{m-1}  b_j \<k^j\> - \sum_{j=1}^{m-1}  b_j \<k^j\> = \<k^m\>. \nonumber
\end{eqnarray}

For example, the 2-qubit observable $O_2$ giving us the second moment of the global distribution $q_k$ reads
\begin{eqnarray*}
    O_2 = n(n-1)|11\>\<11| + O_1\otimes I_1  =n|10\>\<10|+n^2|11\>\<11| .
\end{eqnarray*}
Thus, any pair of nodes performing a collective two-qubit measurement can access both the mean excitation number and additionally also the variance of the global observable.

Similarly, for the 3rd moment of the distribution $q_k$ we have 3-qubit observable
\begin{eqnarray}
    O_3 &=& n(n-1)(n-2)|111\>\<111| + 3O_2\otimes I_1 - 2 O_1\otimes I_2 \nonumber \\
     &=& n|001\>\<001| + n(3n-1) |011\>\<011| + n^3 |111\>\<111|. \nonumber
\end{eqnarray}
We point out that the observables $O_m$ obtained in this manner do not form a unique solution. For example, any 3-qubit observable (\ref{mqubit_general_sigma_average}) yields the same expectation value provided that its coefficients satisfy $T_0=\sigma_{000}=0$, $T_1=\sigma_{001}+\sigma_{010}+\sigma_{100}=n$, $T_2=\sigma_{011}+\sigma_{110}+\sigma_{101}=n(3n-1)$, $T_3=\sigma_{111}=n^3$, while all its off-diagonal elements can be chosen arbitrarily. Analogously, once at least one $m$-qubit observable (\ref{mqubit_general_sigma_average}) has been constructed, all other such observables can be generated. Specifically, they must preserve the same obtained parameters $T_{i}$.

\subsection{Size of the network}
\label{sec:size of the network}

In the previous sections, we assumed that the size of the network is known, which allows the nodes to select observables whose mean values yield the $m$-moments of the probability distribution (\ref{probability_distribution_of_k_particles}). However, one may instead consider a scenario in which the network size is unknown. In that case, individual nodes can independently determine this information through local measurements. We emphasize that this does not contradict the preservation of information about the network state discussed in the previous section. The size information concerns the structure of the network itself, not its state. 

There is a rather straightforward way to get the network size estimation. We can inject a single excitation into an arbitrarily chosen node of the network. The evolution of the initial state $|\psi_{in}\> = |1,0,\ldots ,0\>$ brings it into the asymptotic state
\begin{equation}
    \rho_{as}= \frac{1}{n}P_1,
\end{equation}
whose one-qubit reduced state reads in computational basis
\begin{eqnarray}
    \rho_1=\left(\begin{array}{cc}
       \frac{n-1}{n}  & 0 \\
       0  &  \frac{1}{n}
    \end{array}\right). \nonumber
\end{eqnarray}
Suppose that the first node intends to find out the size of the network. By performing local measurement 
\begin{equation}
    M_1=|0\>\<0|\otimes I_{n-1}, \quad M_2=|1\>\<1|\otimes I_{n-1}.
\end{equation}
the state $|0\>$ is obtained with probability $r_0=(n-1)/n$ and the total state after measurement is given as
\begin{equation}
    \rho^{(0)}=\frac{1}{n-1} |0\>\<0| \otimes P_1^{(n-1)},
\end{equation}
where $P_1^{(n-1)}$ is the operator (\ref{def_projections}) but acting on $n-1$ qubits. Similarly, the state $|1\>$ is measured with probability $r_1=1/n$ and the total state after measurement is $|100\ldots 0\>$. The size of the network is equal to $n=r_0/r_1 + 1$. Moreover, the overall state of the network is not affected by the measurement for all the other nodes of the network. Indeed, the overall state of the network after the measurement is given as
\begin{eqnarray}
    \rho^{,} = r_0 \rho^{(0)} + r_1 \rho^{(1)}=  \frac{1}{n} P_1. \nonumber
\end{eqnarray}

\section{Conclusion}
\label{sec:conclusion}


In this work, we studied quantum consensus in qudit networks driven by discrete random unitary dynamics from the perspective of information redistribution. Using the attractor formalism, we characterized the asymptotic behavior of such networks, identified the global information preserved by the evolution, and showed how it can be recovered from local measurements. This interpretation views quantum consensus not only as agreement among network nodes but also as a mechanism for encoding globally distributed information into locally accessible observables.


We further proved a general theorem specifying which global properties of a multipartite quantum state remain recoverable from expectation values of arbitrary local observables in the asymptotic regime. The result provides a systematic framework for analyzing information accessibility in random unitary networks beyond consensus states.


As applications, we showed that strongly connected CNOT networks with more than two nodes asymptotically reach symmetric state consensus and determined the global information accessible to individual nodes and cooperating groups of nodes. We also solved the inverse problem for average-consensus dynamics by identifying the complete class of two-qubit interactions that asymptotically achieve consensus on the average excitation number, and demonstrated that cooperation among $k$ nodes enables recovery of the first $k$ moments of the excitation-number distribution.

More broadly, our results establish a direct connection between the asymptotic dynamics of random unitary networks and the information that remains experimentally accessible through local measurements. We believe that this perspective provides a useful framework for the analysis and design of quantum consensus protocols, as well as for broader tasks in distributed quantum information processing, where global information must be inferred from local observations. 

Important directions for future research include characterizing the full range of recoverable global properties, particularly in entangled consensus states, and extending our results to non-unital dynamics, which would substantially enlarge the class of admissible consensus protocols and enable a systematic study of information redistribution in dissipative quantum networks.

\section{Acknowledgment}
The authors want to express their gratitude to Kim Nguyen, Oliver Muth, and J\"org Fischer for their continuous encouragement and support.

\bibliographystyle{IEEEtran}
\bibliography{library}

\end{document}